\documentclass[aps,prl,twocolumn,amsmath,amssymb,superscriptaddress,longbibliography]{revtex4-1}
\usepackage{graphicx}
\usepackage{dcolumn}
\usepackage{bm}
\usepackage{amsmath}
\usepackage{color}

\begin{document}

\title{
Electronic signature of the instantaneous asymmetry in the first coordination shell of liquid water
}

\author{Thomas D. K\"uhne}
\affiliation{
Institute of Physical Chemistry, Johannes Gutenberg University Mainz, Staudingerweg 7, 55128 Mainz, Germany
}
\affiliation{
Center for Computational Sciences, Johannes Gutenberg University Mainz, 55128 Mainz, Germany
}
\author{Rustam Z. Khaliullin}
\email{rustam@khaliullin.com}
\affiliation{
Institute of Physical Chemistry, Johannes Gutenberg University Mainz, Staudingerweg 7, 55128 Mainz, Germany
}
\date{\today}

\begin{abstract}
Interpretation of the X-ray spectra of water as evidence for its asymmetric structure has challenged the conventional symmetric nearly-tetrahedral model and initiated an intense debate about the order and symmetry of the hydrogen bond network in water. Here, we present new insights into the nature of local interactions in water obtained using a novel energy decomposition method. Our simulations reveal that while a water molecule forms, on average, two strong donor and two strong acceptor bonds, there is a significant asymmetry in the energy of these contacts. We demonstrate that this asymmetry is a result of small instantaneous distortions of hydrogen bonds, which appear as fluctuations on a timescale of hundreds of femtoseconds around the average symmetric structure. Furthermore, we show that the distinct features of the X-ray absorption spectra originate from molecules with high instantaneous asymmetry. Our findings have important implications as they help reconcile the symmetric and asymmetric views on the structure of water.
\end{abstract}

\maketitle

A detailed description of the structure and dynamics of the hydrogen bond network in water is essential to understand the unique properties of this ubiquitous and important liquid. It has long been accepted that the local structure of water at ambient conditions is tetrahedral~\cite{a:bernal-fowler,a:stillinger-science-1980,a:saykally-water-review-2010}. Although the thermal motion causes distortions from the perfectly tetrahedral configuration, each molecule in the liquid is bonded, on average, to four nearest neighbours via two donor and two acceptor bonds~\cite{a:stillinger-science-1980}. This traditional view is based on results from X-ray and neutron diffraction experiments~\cite{a:soper-water-xray,a:thg-xray-water-review,a:thg-xray-water-2003}, vibrational spectroscopy~\cite{a:tokmakoff-geissler-2003,a:geissler-tokmakoff-2005,a:unified-saykally-2005,a:skinner-review-2010}, macroscopic thermodynamics data~\cite{b:water1969,a:stillinger-science-1980,a:saykally-water-review-2010} as well as molecular dynamics simulations~\cite{a:thg-xray-water-review,a:hutter-water-hybrid-dft,a:tuckerman-water-2006,a:voth-water-review-2009,a:kuhnewater,a:saykally-water-review-2010}.

But this traditional picture has recently been questioned based on data from the X-ray absorption, X-ray emission and X-ray Raman scattering experiments~\cite{a:nilsson,a:water-xes-1,a:water-xes-2,a:water-xes-3}. The results of these spectroscopic studies have been interpreted as evidence for strong distortions in the hydrogen bond network with highly asymmetric distribution of water molecules around a central molecule. It has been suggested that a large fraction of molecules form only two strong hydrogen bonds: one acceptor and one donor bond~\cite{a:nilsson,a:simulation-xray-nilsson,a:water-xes-1,a:water-xes-2,a:water-xes-3,a:nilsson-pettersson-perspective}. However, the``rings and chains'' structure of liquid water~\cite{a:nilsson}, as well as the inhomogeneous two-state model~\cite{a:water-xes-1,a:water-saxs-1} implied by such an interpretation, have been challenged on many fronts~\cite{a:saykally0,a:unified-saykally-2005,a:saykally1,a:thg,a:thg1,a:water-saxs-thg,a:xas-prendergast-galli,a:water-artacho,a:saykally-water-review-2010,a:soper-myths,a:car-xas-water} and are a matter of an ongoing debate~\cite{a:saykally0,a:nilsson-comment1,a:saykally-response1,a:unified-saykally-2005,a:soper,a:saykally1,a:water-saxs-1,a:xas-prendergast-galli,a:water-artacho,a:thg,a:thg1,a:simulation-xray-nilsson,a:water-chains-rmc,a:one-more-nilsson,a:water-saxs-thg,a:water-models-2009,a:nilsson-pettersson-perspective,a:saykally-water-review-2010,a:soper-myths,a:car-xas-water}.

In this work, we performed an unprecedented computational study of the energetics and symmetry of local interactions between water molecules in the liquid phase by utilizing a novel energy decomposition analysis based on absolutely localized molecular orbitals (ALMO EDA)~\cite{a:theeda}. The decomposition of the interaction energy into physically meaningful components provides a deeper insight into the nature and mechanisms of intermolecular bonding than the traditional total-energy electronic structure methods~\cite{a:km,a:rvsX,a:csov1,a:nedaDFT,a:blweda,a:theeda,a:cta}. The success of ALMO EDA in investigating chemical bonding in molecular gas-phase complexes including small water clusters~\cite{a:khalh2o,a:app_almo_2,a:app_almo_4} has inspired us to generalize this methodology and develop the first energy decomposition method for periodic condensed phase systems. 

ALMO EDA separates the total interaction energy of molecules ($\Delta E_{TOT}$) into the interaction energy of the unrelaxed electron densities on the molecules ($\Delta E_{FRZ}$) and the orbital relaxation energy. The latter can be further decomposed into an intramolecular relaxation associated with polarization of the electron clouds on molecules in the field of each other ($\Delta E_{POL}$), two-body donor-acceptor orbital interactions ($\Delta E_{DEL}$) and a generally small higher-order ($\Delta E_{HO}$) relaxation term (see Ref.~\onlinecite{a:theeda} for a detailed description of the ALMO EDA terms).
\begin{eqnarray}
\label{eq:etot}
\Delta E_{TOT} & = & \Delta E_{FRZ} + \Delta E_{POL} + \Delta E_{DEL} + \Delta E_{HO}\nonumber\\
\label{eq:edel}
\Delta E_{DEL} & = & \sum_{C = 1}^{Mol} \Delta E_{C} = \sum_{D,A = 1}^{Mol} \Delta E_{D \rightarrow A}
\end{eqnarray}

Two-body components $\Delta E_{D \rightarrow A}$ are the main focus of this work and arise due to delocalization of electrons (or charge transfer) from the occupied orbitals of donor molecule $D$ to the virtual orbitals of acceptor $A$. Each of these terms provide an accurate measure of the perturbation of orbitals localized on a molecule by donor or acceptor interactions with a single neighbour in the many-body system. The donor-acceptor energies are obtained self-consistently and include cooperativity effects, which are essential for a correct description of the hydrogen bond networks~\cite{a:water-cooperativity0,a:nedaDFT}. Furthermore, the two-body terms are natural local descriptors of intermolecular bonding, which allowed us to analyze the molecular network in liquid water without introducing any arbitrary definitions of a hydrogen bond~\cite{a:kumar}.

The unique insight obtained by applying ALMO EDA to liquid water revealed a significant asymmetry in the strength of the local donor-acceptor contacts. \emph{Ab initio} molecular dynamics (AIMD) simulations performed with the recently developed second-generation Car-Parrinello approach~\cite{a:2ndcpmd} enabled us to characterize geometric origins of the asymmetry, its dynamical behaviour, as well as the mechanism of its relaxation. Furthermore, to address the controversial question of whether it is correct to interpret the X-ray spectra of water in terms of asymmetric structures we performed extensive calculations of its X-ray absorption (XA) spectrum and compared the spectral characteristics of water molecules with different degree of asymmetry. 

\section{Results}

\textbf{Electronic asymmetry and its origins.} ALMO EDA was performed for the decorrelated configurations collected from an AIMD simulation at constant temperature (300~K) and density (0.9966~g/cm$^3$, see Methods for computational details). The electron delocalization energy per molecule $\Delta E_{C}$ defined in Equation (\ref{eq:edel}) was analysed by considering each water molecule as a donor and as an acceptor:
\begin{eqnarray}
\label{eq:edelmol}
\Delta E_{C} & = & \sum_{N=1}^{Mol} \Delta E_{C \rightarrow N} = \sum_{N=1}^{Mol} \Delta E_{N \rightarrow C}, \nonumber
\end{eqnarray}
where $C$ stands for the central molecule and $N$ for its neighbours. It is important to emphasize that terms \emph{donor} and \emph{acceptor} are used throughout this paper to describe the role of a molecule in the transfer of the electron density. This is opposite to the labeling used for a donor and an acceptor of hydrogen in a hydrogen bond.

\begin{figure}
\includegraphics*[width=8cm]{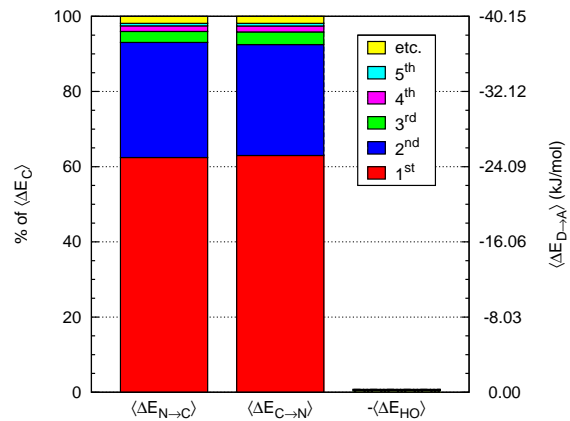}
\caption{\label{fig:del5} \textbf{Five strongest interactions.} Average contributions of the strongest acceptor $\langle \Delta E_{N \rightarrow C} \rangle$ and donor $\langle \Delta E_{C \rightarrow N} \rangle$ interactions. The rightmost column shows that higher-order delocalization $\langle \Delta E_{HO} \rangle$ does not contribute to the overall binding significantly.}
\end{figure}

Figure~\ref{fig:del5} shows contributions of the five strongest donor-acceptor interactions to the average delocalization energy of a molecule $\langle \Delta E_{C}\rangle$ (brackets $\langle ..\rangle$ denote averaging over all central molecules and AIMD snapshots). It demonstrates that electron delocalization is dominated by two strong interactions, which together are responsible for $\sim$93\% of the delocalization energy of a single molecule. The third and the fourth strongest donor (acceptor) interactions contribute $\sim$5\% and correspond to back-donation of electrons to (from) the remaining two first-shell neighbours (i.e. there is non-negligible delocalization from a typical acceptor to a typical donor). The remaining $\sim$2\% correspond to the delocalization energy to (from) the second and more distant coordination shells.

\begin{figure}
\includegraphics*[width=8cm]{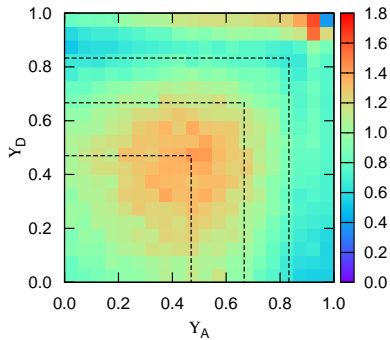}
\caption{\label{fig:asymhist} \textbf{The normalized probability density function of the asymmetry parameters $\Upsilon_{A}$ and $\Upsilon_{D}$.} The probability of finding a molecule in a bin can be found by dividing the corresponding density value by the number of bins (i.e. 400). The dashed black lines at $\Upsilon \approx \frac{1}{2},\frac{2}{3},\frac{5}{6}$ partition all molecules into four groups of equal sizes.}
\end{figure}

Comparison of the strengths of the first and second strongest donor-acceptor interactions ($\sim$25~kJ/mol and $\sim$12~kJ/mol, respectively) with that in water dimer ($\sim$9~kJ/mol) suggests that each water molecule can be considered to form, on average, two donor and two acceptor bonds. Substantial difference in the strengths of the first and second strongest interactions implies that a large fraction of water molecules experience a significant asymmetry in their local environment. To characterize the asymmetry of the two strongest donor contacts of a molecule we introduced a dimensionless asymmetry parameter
\begin{eqnarray}
\label{eq:asym}
\Upsilon_{D} = 1 - \frac{\Delta E_{C \rightarrow N^{2nd}}}{\Delta E_{C \rightarrow N^{1st}}} \nonumber
\end{eqnarray}
and an equivalent parameter for the acceptor interactions $\Upsilon_{A}$. The asymmetry parameter is zero if the two contacts are equally strong and equals to one if the second contact does not exist. The probability distribution of molecules according to their $\Upsilon$-parameters is shown in Figure~\ref{fig:asymhist} together with the lines separating the molecules into four groups of equal sizes with different asymmetry. The shape of the distribution demonstrates that most molecules form highly asymmetric donor or acceptor contacts at any instance of time. For example, the line at $\Upsilon \approx 0.5$ indicates that for $\sim$75\% of molecules either $\Upsilon_{A}$ or $\Upsilon_{D}$ is more than 0.5, which means that the strongest donor or acceptor contact is at least two times stronger than the second strongest for these molecules. Furthermore, the peak in the region of high $\Upsilon$ in Figure~\ref{fig:asymhist} indicates the presence of molecules with significantly distorted hydrogen bonds.

\begin{figure}
\includegraphics*[width=8.5cm]{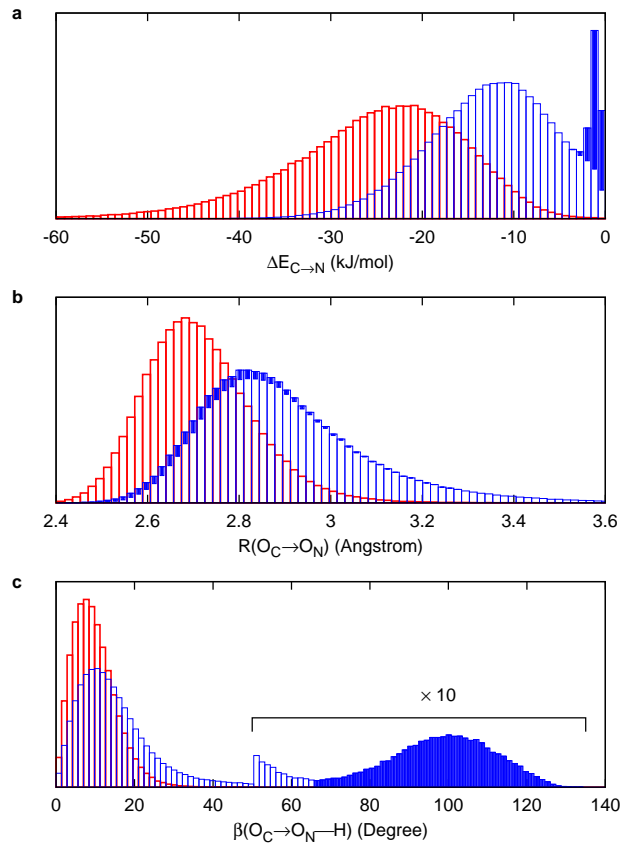}
\caption{\label{fig:hist0} \textbf{Energetic and geometric characteristics of the instantaneous asymmetry.} The probability distribution of the (\textbf{a}) strength, (\textbf{b}) intermolecular distance $R$ and (\textbf{c}) hydrogen bond angle $\beta$ for the first (red) and second (blue) strongest donor interactions $C \rightarrow N$. Filled areas show the contribution of configurations, for which back-donation to a nearby donor is stronger than donation to the second acceptor (cf. text). Distributions for acceptor interactions, $N \rightarrow C$, are almost identical and not shown.}
\end{figure}

To understand the origins of the asymmetry we compared the geometry of donor-acceptor pairs involved in the first and second strongest interactions. We found that the strength of the interaction is greatly affected by the intermolecular distance $R \equiv d(O_{D} - O_{A})$ and the hydrogen bond angle $\beta \equiv \angle O_{D}O_{A}H$ while the other geometric parameters have only minor influence on $\Delta E_{D\rightarrow A}$. The strongly overlapping distributions in Figure~\ref{fig:hist0} suggest that some second strongest interactions have the same energetic and geometric characteristics as the strongest contacts. This implies that the observed electronic asymmetry cannot be attributed to the presence of two distinct types of hydrogen bonds -- weak and strong. It is rather a result of continuous deformations of a typical bond. Another important conclusion that can be made from the distributions in Figure~\ref{fig:hist0} is that relatively small variations of the intermolecular distance ($R\sim 0.2~\AA$) and hydrogen bond angle ($\beta\sim 5-10^{\circ}$) lead to the remarkable changes in the strength and electronic structure of hydrogen bonds. Analysis of the structure of the molecular chains defined by the first strongest bonds (i.e. one donor and one acceptor for each molecule) shows that their directions are random, without any long-range order (i.e. rings, spirals or zig-zags) on the length scale of the simulation box ($\sim15~\AA$).
 
We did not attempt to quantify the concentration of single-donor and single-acceptor molecules in this work because the slow decay of the distribution tails (Figure~\ref{fig:hist0}) implies that defining such configurations using a distance, angle or energy cutoff is an unavoidably arbitrary procedure. A quantitative analysis of the network, which was performed for the same molecular configurations in Ref.~\onlinecite{a:kuhnewater}, shows that according to the commonly used geometric definitions of hydrogen bonds~\cite{a:luzar,a:aimd3,a:nilsson} the structure of water is distorted tetrahedral with only a small fraction of broken bonds. The results presented in Figure~\ref{fig:hist0} indicate that these geometric criteria cannot fully characterize the dramatic effect of distortions on the local electronic structure and donor-acceptor interactions of water molecules.  

It is important to note that some second strongest interactions are weakened by distortions to such an extent that back-donation to (from) a nearby donor (acceptor) becomes the second strongest interaction. Such configurations can be clearly distinguished by the large-angle peak in Figure~\ref{fig:hist0}c (the region of the 10-fold magnification). They account for $\sim 6-7$\% of all configurations and are responsible for the low-energy peak in the distribution of $\Delta E_{D \rightarrow A}$ (filled blue areas in Figure~\ref{fig:hist0}) and for the high-$\Upsilon$ peak in Figure~\ref{fig:asymhist}.

\begin{figure}
\includegraphics*[width=8.5cm]{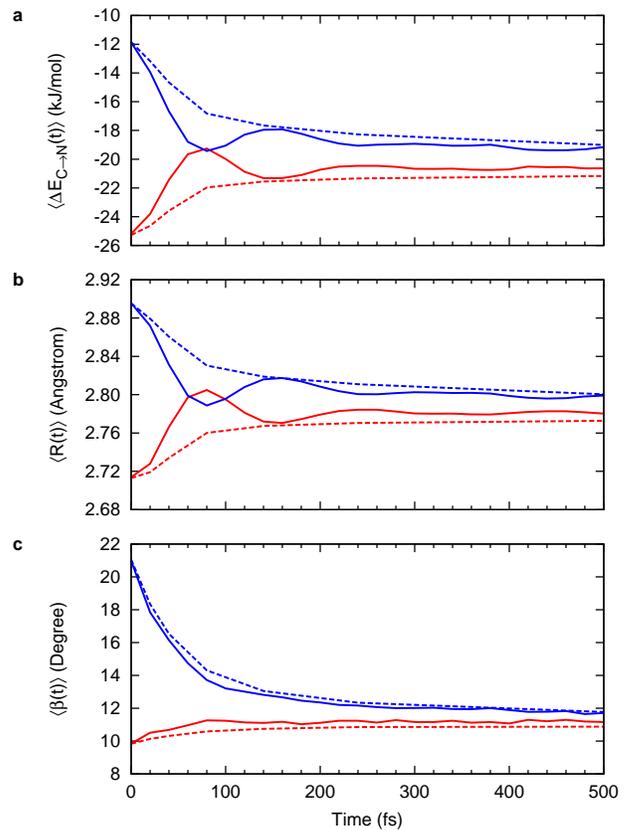}
\caption{\label{fig:trelax} \textbf{Relaxation of the instantaneous asymmetry.} Time dependence of the (\textbf{a}) average strength, (\textbf{b}) intermolecular distance $R$ and (\textbf{c}) hydrogen bond angle $\beta$ for the first (red) and second (blue) strongest donor interactions $C \rightarrow N$. Solid lines show the instantaneous values while the dashed lines correspond to the time-average values (see Methods). Time-dependent characteristics of acceptor interactions, $N \rightarrow C$, are almost identical and not shown.}
\end{figure}

\textbf{Relaxation of the instantaneous asymmetry.} The overlapping distributions in Figure~\ref{fig:hist0} suggest that despite the difference in the strength of the donor-acceptor contacts their nature is similar and in the process of thermal motion the strongest interacting pair can become the second strongest and \emph{vice versa}. To estimate the time scale of this process, we examined how the average energy of the first two strongest interactions fluctuates in time (see Methods for definitions of time-dependent averages). Figure~\ref{fig:trelax}a shows that the strength of these interactions oscillates rapidly and after $\sim$80~fs from an arbitrarily chosen time origin, the first strongest interaction becomes slightly weaker than the second strongest (note that \emph{first} and \emph{second} refer to their order at $t = 0$). The amplitude of the oscillations decreases and the strengths of both interactions approach the average value of $\sim$20~kJ/mol on the time scale of hundreds of femtoseconds. The decay of the oscillations indicates fast decorrelation of the time-separated instantaneous values because of the strong coupling of a selected pair of molecules with its surroundings. In other words, although the energy of a particular hydrogen bond fluctuates around its average value indefinitely (i.e. with a never-decreasing amplitude), this bond has approximately equal chances of becoming weak or strong after a certain period of time independently of its strength at $t = 0$. This effect is due to the noise introduced by the environment and can be observed in ultrafast infrared spectroscopy experiments~\cite{a:tokmakoff-geissler-2003}.

It is important to emphasize that the time averages shown in Figure~\ref{fig:trelax} are physically meaningful and can be calculated accurately only for the time intervals that are shorter than the average lifetime of a hydrogen bond $\tau_{\text{HB}} \approx 5$~ps~\cite{a:luzar,a:kuhnewater} (see Methods). The small residual asymmetry that is still present after 500~fs (Figure~\ref{fig:trelax}a) is an indication of the slow non-exponential relaxation behaviour that characterises the kinetics of many processes in liquid water~\cite{a:luzar}.

In addition to the instantaneous values of $\Delta E_{D\rightarrow A}$ \emph{at} time $t$, the dashed lines in Figure~\ref{fig:trelax}a show the corresponding averages \emph{over} time $t$ (see Methods). The latter shows that any neighbour-induced asymmetry in the electronic structure of a water molecule can be observed only with an experimental probe with a time-resolution of tens of femtoseconds or less. On longer time scales the asymmetry is destroyed by the thermal motion of molecules and only the average symmetric structures can be observed in experiments with low temporal resolution. 

An examination of time dependence of all two-body and some three-body geometric parameters that characterize the relative motion of molecules reveals the mechanism of the relaxation. Similar shapes of the curves in Figures~\ref{fig:trelax}a and \ref{fig:trelax}b shows that the relaxation of the asymmetry is primarily caused by low-frequency vibrations of the molecules relative to each other. The minor differences in the behaviour of the curves, in particular at 80~fs, indicate that the relaxation of the asymmetry is also influenced by some other degrees of freedom. The temporal changes in the hydrogen bond angles towards the average value (Figure~\ref{fig:trelax}c) show that librations of molecules play this minor role in the relaxation process. The absence of the correlation between the energy curves and the remaining geometric parameters indicates that the other types of motion are not important for the relaxation mechanism. 

The kinetics and mechanism of the asymmetry relaxation presented here are supported by data from ultrafast infrared spectroscopy, which can directly observe intermolecular oscillations with a period of 170~fs~\cite{a:tokmakoff-geissler-2003}. They are also in agreement with the theoretical work of Artacho \emph{et al.}, who have used the Mulliken bond order parameter to characterise the connectivity and dynamical processes in the hydrogen bond network of liquid water~\cite{a:water-artacho}. 

\begin{figure}
\includegraphics*[width=8.5cm]{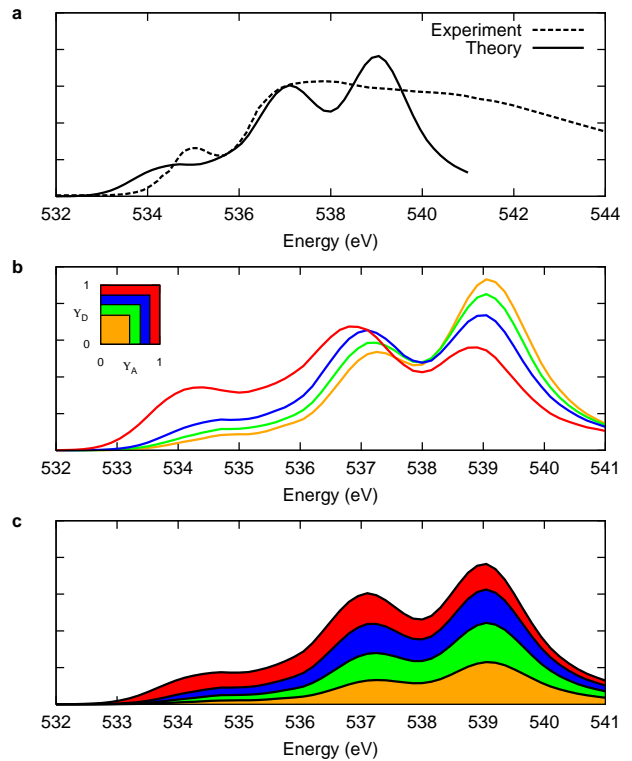}
\caption{\label{fig:xas} \textbf{X-ray absorption (XA) spectra of liquid water.} (\textbf{a}) Comparison of the calculated and experimental~\cite{a:nilsson} XA spectra. (\textbf{b}) Calculated XA spectra of the four groups of molecules separated according to the asymmetry of their donor ($\Upsilon_{D}$) and acceptor ($\Upsilon_{A}$) environments as shown in the inset. (\textbf{c}) Contributions of the four groups into the total XA spectrum. The color coding is shown in the inset above.}
\end{figure}

\textbf{X-ray absorption signatures of asymmetric structures.} The time behaviour described above implies that the instantaneous asymmetry can, in principle, be detected by X-ray spectroscopy, which has temporal resolution of several femtoseconds and is highly sensitive to perturbations in the electronic structure of molecules~\cite{a:nilsson,a:nilsson-pettersson-perspective}. To identify possible relationships between the spectroscopic features and asymmetry, we calculated the XA spectrum of liquid water at the oxygen K-edge using the half-core-hole transition potential formalism~\cite{a:hch-water-review} within all-electron density functional theory~\cite{a:xas-hutter} (see Methods for computational details). Although the employed computational approach overestimates intensities in the post-edge part of the spectrum and underestimates the pre-edge peak and overall spectral width~\cite{a:xas-iannuzzi} it provides an accurate description of the core-level excitation processes and semiquantitatively reproduces the main features of the experimentally measured spectra (Figure~\ref{fig:xas}a). 

The localized nature of the $1s$ core orbitals allows to disentangle spectral contributions from molecules with different asymmetry. To this end, all molecules were separated into four groups according to the asymmetry of their donor and acceptor environments as shown in Figure~\ref{fig:asymhist}. Choosing boundaries at $\Upsilon \approx \frac{1}{2},\frac{2}{3},\frac{5}{6}$ distributes all molecules into four groups of approximately equal sizes (i.e. $25\pm 2\%$). Figure~\ref{fig:xas}b shows four XA spectra obtained by averaging the individual contributions of molecules in each group. It reveals that molecules in the symmetric environment exhibit strong post-edge peaks while molecules with high asymmetry of their environment are characterized by the amplified absorption in the pre-edge region. Furthermore, the relationship between the asymmetry and absorption intensity is nonuniform: the pre-edge peak is dramatically increased in the spectrum for the 25\% of molecules in the most asymmetric group, for which the first strongest interaction is more than six times stronger than the second. As a consequence, the pre-edge feature of the calculated XA is dominated by the contribution of molecules in the highly asymmetric environments (Figure~\ref{fig:xas}c).

The pronounced pre-edge peak in the experimentally measured XA spectrum of liquid water has been interpreted as evidence for its ``rings and chains'' structure, where $\sim$80\% of molecules have two broken hydrogen bonds~\cite{a:nilsson,a:simulation-xray-nilsson}. Our results suggest that this feature of the XA spectrum can be explained by the presence of a smaller fraction of water molecules with high instantaneous asymmetry. Although the employed XA modeling methodology does not allow us to estimate precisely the size of this fraction our result is consistent with that of recent theoretical studies at an even higher level of theory, which have demonstrated that the main features of the experimental XA spectra can be reproduced in simulations based on conventional nearly-tetrahedral models~\cite{a:car-xas-water,a:kong-xas}. Our work complements the previous results by revealing an interesting and important connection between relatively small geometric perturbations in the hydrogen-bond network, the large asymmetry in the electronic ground state and the XA spectral signatures of the core-excitation processes.

\textbf{Conclusions.} We have applied a combination of ALMO EDA, XA spectra modelling and second-generation Car-Parrinello molecular dynamics, to study perturbation induced in the electronic structure of water molecules by local donor-acceptor interactions with their neighbours in the liquid phase. Our main conclusions are as follows. First, the strength of donor-acceptor interactions suggest that each molecule in liquid water at ambient conditions forms, on average, two donor and two acceptor bonds. Second, even small thermal distortions in the tetrahedral hydrogen bond network induce a significant asymmetry in the strength of the contacts causing one of the two donor (acceptor) interactions to become, at any instance of time, substantially stronger than the other. Thus, the instantaneous structure of water is strongly asymmetric only according to the electronic criteria, not the geometric one. Third, intermolecular vibrations and librations of OH groups of hydrogen bonds result in the relaxation of the instantaneous asymmetry on the time scale of hundreds of femtoseconds. Finally, the pronounced pre-edge peak observed in the XA spectra of liquid water can be attributed to molecules in asymmetric environments created by instantaneous distortions in the fluctuating but on average symmetric hydrogen bond network.

Thus, our work helps reconcile the two existing and seemingly opposite models of liquid water -- the traditional symmetric and the recently proposed asymmetric -- and represents an important step towards a better understanding of the electronic structure of the hydrogen bond network of one of the most important liquids on Earth.

The broader significance of this work lies in the development of an energy decomposition method for condensed phase systems, which can provide deep insight into the electronic structure of a wide range of molecular systems and, thus, offers new opportunities for studying their complex behaviour.

\section{Methods}

\textbf{\emph{Ab initio} molecular dynamics.} \emph{Ab initio} MD simulations with classical nuclei based on the Perdew-Burke-Ernzerhof (PBE) exchange-correlation (XC) functional were performed at constant temperature (300~K) and density (0.9966~g/cm$^3$). Long 70~ps AIMD trajectories for systems containing 128 water molecules were obtained by taking advantage of the computational efficiency of the second-generation Car-Parrinello method~\cite{a:2ndcpmd} as described in Ref.~\onlinecite{a:kuhnewater}. Our previous work~\cite{a:kuhnewater} has shown that PBE provides a realistic description of many important structural and dynamical characteristics of liquid water including the radial distribution functions, self-diffusion and viscosity coefficients, vibrational spectrum and hydrogen-bond lifetime. However, the PBE water with the classical description of the nuclei is somewhat overstructured suggesting that the degree of the network distortion, fraction of broken bonds and, thus, the asymmetry in our simulations may be underestimated. Therefore, we verified that the main conclusions presented in this work are also valid for snapshots generated with simulations with quantum hydrogen nuclei and based on a water model, which rectifies the main shortcomings of the PBE functional (see Supplementary Information).


\textbf{Energy decomposition analysis.} ALMO EDA~\cite{a:theeda} for periodic systems was implemented~\cite{a:khal} in the CP2K package, which relies on the mixed Gaussian and plane wave (GPW) representation of the electronic degrees of freedom~\cite{a:quickstep}. The GPW approach makes CP2K uniquely suited for ALMO EDA since the localized atom-centered Gaussian basis sets are required for the construction of absolutely localized molecular orbitals, whereas the use of plane waves ensures computational efficiency of the large-scale calculations performed in this work.

Molecular orbitals in all ALMO EDA calculations were represented by a triple-$\zeta$ Gaussian basis set with two sets of polarization functions (TZV2P) optimized specifically for molecular systems~\cite{a:molopt}. A high density cutoff of 1000 Ry was used to describe the electron density. The XC energy was approximated with the BLYP functional. The Brillouin zone was sampled at the $\Gamma$-point and separable norm-conserving pseudopotentials were used to describe the interactions between the valence electrons and the ionic cores~\cite{a:hgh}. We verified that performing ALMO EDA with the HSE06 screened hybrid XC functional, which provides better description of band gaps and electron delocalization effects than BLYP, does not change the main conclusions of the work (see Supplementary Information).

\textbf{Averaging procedures.} The average donor-acceptor energies shown in Figure~\ref{fig:del5} and distributions in Figures~\ref{fig:asymhist} and \ref{fig:hist0} were obtained by performing ALMO EDA for 701 decorrelated AIMD snapshots separated by 100~fs (i.e. 89,728 molecular configurations).

The time-dependent averages in Figure~\ref{fig:trelax}a were computed using the ALMO EDA terms for 3,501 snapshots separated by 20~fs (448,128 local configurations). The instantaneous values at time $t$ (solid lines in Figure~\ref{fig:trelax}a) were calculated by averaging over time origins $\tau$ separated by 100~fs and over all \emph{surviving triples}:
\begin{eqnarray}
\label{eq:atime}
\langle \Delta E_{C\rightarrow N}(t) \rangle & = & \frac{1}{T} \sum_{\tau=1}^{T} \frac{1}{M(\tau,t)} \sum_{C=1}^{M(\tau,t)} \Delta E_{C \rightarrow N}(\tau+t), \nonumber \end{eqnarray}
where $M(\tau,t)$ is the number of triples that survived from time $\tau$ to $\tau + t$. A triple is considered to survive a specified time interval if the central molecule has the same two strongest-interacting neighbours in all snapshots in this interval.

The average values over time $t$ (dashed lines in Figure~\ref{fig:trelax}a) were calculated by averaging over time origins $\tau$, all snapshots lying in the time interval from $\tau$ to $\tau + t$ and over all surviving triples:
\begin{eqnarray}
\label{eq:aovertime}
\langle &\Delta &E_{C \rightarrow N} (t) \rangle^{*} = \nonumber \\
&=& \frac{1}{T} \sum_{\tau=1}^{T} \frac{1}{t+1} \sum_{\kappa=0}^{t} \frac{1}{M(\tau,t)} \sum_{C=1}^{M(\tau,t)} \Delta E_{C \rightarrow N}(\tau+\kappa) \nonumber
\end{eqnarray}
Time-averages $\langle A(t) \rangle^{*}$ are related to the instantaneous values $\langle A(t) \rangle$ by the following equation:
\begin{eqnarray}
\label{eq:reation}
\langle A(t) \rangle^{*} & \approx & \frac{1}{t} \int_{0}^{t} \langle A(\tau) \rangle d\tau, \nonumber
\end{eqnarray}
where equality holds if all triples survive over time $t$.

The time averages defined here are physically meaningful and can be accurately calculated only for the time intervals that are shorter than the average lifetime of a triple or, equivalently, the average lifetime of a hydrogen bond, $\tau_{\text{HB}} \approx 5$~ps~\cite{a:luzar,a:kuhnewater}. Nevertheless, some triples break apart even after shorter time affecting the time averages. For example, only $\sim$80\% of all triples survive over the time interval shown in Figure~\ref{fig:trelax}. We found that stronger interacting triples have longer lifetimes and, therefore, are overrepresented at long time. However, the error introduced by this bias is only $\sim$1.5~kJ/mol at 0.5~ps and does not affect the main conclusions. This error is the main reason that the long-time averages in Figure~\ref{fig:trelax}a converge to 20~kJ/mol instead of the expected value of 18.5~kJ/mol. 


\textbf{X-ray absorption spectrum.} The XA spectra of water at the oxygen K-edge were obtained using the all-electron Gaussian augmented plane wave formalism implemented in the CP2K package~\cite{a:xas-hutter,a:xas-iannuzzi}. The calculations were performed with the BLYP XC functional and half-core-hole transition potential. Large basis sets (6-311G** for hydrogen and cc-pVQZ for oxygen atoms) were used to provide an adequate representation of the unoccupied molecular orbitals in the vicinity of absorbing atoms~\cite{a:xas-iannuzzi}. A density cutoff of 320 Ry was used to describe the soft part of the electron density. The onset energies of the absorption spectra were aligned with the first $\Delta$SCF excitation energy obtained for each oxygen atom from a separate calculation with the same setup. The spectral intensities were calculated as transition moment integrals in the velocity form. To mimic the experimental broadening, the discrete lines were convoluted with Gaussian functions of 0.2~eV width at half-maximum. The final spectrum was obtained by averaging the convoluted spectra of 9,024 oxygen atoms from 141 AIMD snapshots separated by 500~fs (i.e. 64 oxygen atoms were randomly selected in each snapshot). 

\section{Acknowledgments}

Dedicated to Alexis T. Bell on his 70th birthday for the invaluable contribution to the development of ALMO EDA. The authors thank J\"urg Hutter and Joost VandeVondele for their assistance in implementing ALMO EDA in the CP2K package and Thomas Beardsley for critical reading of the manuscript. R.Z.K. is grateful to the Swiss National Science Foundation for financial support and to the Swiss National Supercomputing Centre (CSCS) for computer time. T.D.K. acknowledges financial support from the Graduate School of Excellence MAINZ and the Carl Zeiss Foundation.

\section{Author contributions}

T.D.K. conceived the study, performed molecular dynamics simulations, discussed the results, commented on the manuscript; R.Z.K. conceived the study, implemented the ALMO EDA code, performed ALMO EDA calculations, analyzed and interpreted the results, wrote the manuscript and supervised the project.

\section{Competing ﬁnancial interest}

The authors declare no competing financial interests.


\begin{thebibliography}{60}%
\makeatletter
\providecommand \@ifxundefined [1]{%
 \@ifx{#1\undefined}
}%
\providecommand \@ifnum [1]{%
 \ifnum #1\expandafter \@firstoftwo
 \else \expandafter \@secondoftwo
 \fi
}%
\providecommand \@ifx [1]{%
 \ifx #1\expandafter \@firstoftwo
 \else \expandafter \@secondoftwo
 \fi
}%
\providecommand \natexlab [1]{#1}%
\providecommand \enquote  [1]{``#1''}%
\providecommand \bibnamefont  [1]{#1}%
\providecommand \bibfnamefont [1]{#1}%
\providecommand \citenamefont [1]{#1}%
\providecommand \href@noop [0]{\@secondoftwo}%
\providecommand \href [0]{\begingroup \@sanitize@url \@href}%
\providecommand \@href[1]{\@@startlink{#1}\@@href}%
\providecommand \@@href[1]{\endgroup#1\@@endlink}%
\providecommand \@sanitize@url [0]{\catcode `\\12\catcode `\$12\catcode
  `\&12\catcode `\#12\catcode `\^12\catcode `\_12\catcode `\%12\relax}%
\providecommand \@@startlink[1]{}%
\providecommand \@@endlink[0]{}%
\providecommand \url  [0]{\begingroup\@sanitize@url \@url }%
\providecommand \@url [1]{\endgroup\@href {#1}{\urlprefix }}%
\providecommand \urlprefix  [0]{URL }%
\providecommand \Eprint [0]{\href }%
\@ifxundefined \urlstyle {%
  \providecommand \doi  [0]{\begingroup \@sanitize@url \@doi}%
  \providecommand \@doi [1]{\endgroup \@@startlink {\doibase
  #1}doi:\discretionary {}{}{}#1\@@endlink }%
}{%
  \providecommand \doi  [0]{doi:\discretionary{}{}{}\begingroup
  \urlstyle{rm}\Url }%
}%
\providecommand \doibase [0]{http://dx.doi.org/}%
\providecommand \Doi [0]{\begingroup \@sanitize@url \@Doi }%
\providecommand \@Doi  [1]{\endgroup\@@startlink{\doibase#1}\@@Doi}%
\providecommand \@@Doi [1]{#1\@@endlink}%
\providecommand \selectlanguage [0]{\@gobble}%
\providecommand \bibinfo  [0]{\@secondoftwo}%
\providecommand \bibfield  [0]{\@secondoftwo}%
\providecommand \translation [1]{[#1]}%
\providecommand \BibitemOpen [0]{}%
\providecommand \bibitemStop [0]{}%
\providecommand \bibitemNoStop [0]{.\EOS\space}%
\providecommand \EOS [0]{\spacefactor3000\relax}%
\providecommand \BibitemShut  [1]{\csname bibitem#1\endcsname}%
\bibitem [{\citenamefont {Bernal}\ and\ \citenamefont
  {Fowler}(1933)}]{a:bernal-fowler}%
  \BibitemOpen
  \bibfield  {author} {\bibinfo {author} {\bibfnamefont {J.~D.}\ \bibnamefont
  {Bernal}}\ and\ \bibinfo {author} {\bibfnamefont {R.~H.}\ \bibnamefont
  {Fowler}},\ }\bibfield  {title} {\enquote {\bibinfo {title} {A theory of
  water and ionic solution, with particular reference to hydrogen and hydroxyl
  ions},}\ }\href@noop {} {\bibfield  {journal} {\bibinfo  {journal} {The
  Journal of Chemical Physics},\ }\textbf {\bibinfo {volume} {1}},\ \bibinfo
  {pages} {515--548} (\bibinfo {year} {1933})}\BibitemShut {NoStop}%
\bibitem [{\citenamefont {Stillinger}(1980)}]{a:stillinger-science-1980}%
  \BibitemOpen
  \bibfield  {author} {\bibinfo {author} {\bibfnamefont {F.~H.}\ \bibnamefont
  {Stillinger}},\ }\bibfield  {title} {\enquote {\bibinfo {title} {Water
  revisited},}\ }\href@noop {} {\bibfield  {journal} {\bibinfo  {journal}
  {Science},\ }\textbf {\bibinfo {volume} {209}},\ \bibinfo {pages} {451--457}
  (\bibinfo {year} {1980})}\BibitemShut {NoStop}%
\bibitem [{\citenamefont {Clark}\ \emph
  {et~al.}(2010){\natexlab{a}}\citenamefont {Clark}, \citenamefont {Cappa},
  \citenamefont {Smith}, \citenamefont {Saykally},\ and\ \citenamefont
  {Head-Gordon}}]{a:saykally-water-review-2010}%
  \BibitemOpen
  \bibfield  {author} {\bibinfo {author} {\bibfnamefont {G.~N.~I.}\
  \bibnamefont {Clark}}, \bibinfo {author} {\bibfnamefont {C.~D.}\ \bibnamefont
  {Cappa}}, \bibinfo {author} {\bibfnamefont {J.~D.}\ \bibnamefont {Smith}},
  \bibinfo {author} {\bibfnamefont {R.~J.}\ \bibnamefont {Saykally}}, \ and\
  \bibinfo {author} {\bibfnamefont {T.}~\bibnamefont {Head-Gordon}},\
  }\bibfield  {title} {\enquote {\bibinfo {title} {The structure of ambient
  water},}\ }\href@noop {} {\bibfield  {journal} {\bibinfo  {journal} {Mol.
  Phys.},\ }\textbf {\bibinfo {volume} {108}},\ \bibinfo {pages} {1415--1433}
  (\bibinfo {year} {2010}{\natexlab{a}})}\BibitemShut {NoStop}%
\bibitem [{\citenamefont {Soper}(2000)}]{a:soper-water-xray}%
  \BibitemOpen
  \bibfield  {author} {\bibinfo {author} {\bibfnamefont {A.~K.}\ \bibnamefont
  {Soper}},\ }\bibfield  {title} {\enquote {\bibinfo {title} {The radial
  distribution functions of water and ice from 220 to 673 {K} and at pressures
  up to 400 {MPa}},}\ }\href@noop {} {\bibfield  {journal} {\bibinfo  {journal}
  {Chem. Phys.},\ }\textbf {\bibinfo {volume} {258}},\ \bibinfo {pages}
  {121--137} (\bibinfo {year} {2000})}\BibitemShut {NoStop}%
\bibitem [{\citenamefont {Head-Gordon}\ and\ \citenamefont
  {Hura}(2002)}]{a:thg-xray-water-review}%
  \BibitemOpen
  \bibfield  {author} {\bibinfo {author} {\bibfnamefont {T.}~\bibnamefont
  {Head-Gordon}}\ and\ \bibinfo {author} {\bibfnamefont {G.}~\bibnamefont
  {Hura}},\ }\bibfield  {title} {\enquote {\bibinfo {title} {Water structure
  from scattering experiments and simulation},}\ }\href@noop {} {\bibfield
  {journal} {\bibinfo  {journal} {Chem. Rev.},\ }\textbf {\bibinfo {volume}
  {102}},\ \bibinfo {pages} {2651--2669} (\bibinfo {year} {2002})}\BibitemShut
  {NoStop}%
\bibitem [{\citenamefont {Hura}\ \emph {et~al.}(2003)\citenamefont {Hura},
  \citenamefont {Russo}, \citenamefont {Glaeser}, \citenamefont {Head-Gordon},
  \citenamefont {Krack},\ and\ \citenamefont
  {Parrinello}}]{a:thg-xray-water-2003}%
  \BibitemOpen
  \bibfield  {author} {\bibinfo {author} {\bibfnamefont {G.}~\bibnamefont
  {Hura}}, \bibinfo {author} {\bibfnamefont {D.}~\bibnamefont {Russo}},
  \bibinfo {author} {\bibfnamefont {R.~M.}\ \bibnamefont {Glaeser}}, \bibinfo
  {author} {\bibfnamefont {T.}~\bibnamefont {Head-Gordon}}, \bibinfo {author}
  {\bibfnamefont {M.}~\bibnamefont {Krack}}, \ and\ \bibinfo {author}
  {\bibfnamefont {M.}~\bibnamefont {Parrinello}},\ }\bibfield  {title}
  {\enquote {\bibinfo {title} {Water structure as a function of temperature
  from {X}-ray scattering experiments and ab initio molecular dynamics},}\
  }\href@noop {} {\bibfield  {journal} {\bibinfo  {journal} {Phys. Chem. Chem.
  Phys.},\ }\textbf {\bibinfo {volume} {5}},\ \bibinfo {pages} {1981--1991}
  (\bibinfo {year} {2003})}\BibitemShut {NoStop}%
\bibitem [{\citenamefont {Fecko}\ \emph {et~al.}(2003)\citenamefont {Fecko},
  \citenamefont {Eaves}, \citenamefont {Loparo}, \citenamefont {Tokmakoff},\
  and\ \citenamefont {Geissler}}]{a:tokmakoff-geissler-2003}%
  \BibitemOpen
  \bibfield  {author} {\bibinfo {author} {\bibfnamefont {C.~J.}\ \bibnamefont
  {Fecko}}, \bibinfo {author} {\bibfnamefont {J.~D.}\ \bibnamefont {Eaves}},
  \bibinfo {author} {\bibfnamefont {J.~J.}\ \bibnamefont {Loparo}}, \bibinfo
  {author} {\bibfnamefont {A.}~\bibnamefont {Tokmakoff}}, \ and\ \bibinfo
  {author} {\bibfnamefont {P.~L.}\ \bibnamefont {Geissler}},\ }\bibfield
  {title} {\enquote {\bibinfo {title} {Ultrafast hydrogen-bond dynamics in the
  infrared spectroscopy of water},}\ }\href@noop {} {\bibfield  {journal}
  {\bibinfo  {journal} {Science},\ }\textbf {\bibinfo {volume} {301}},\
  \bibinfo {pages} {1698--1702} (\bibinfo {year} {2003})}\BibitemShut {NoStop}%
\bibitem [{\citenamefont {Eaves}\ \emph {et~al.}(2005)\citenamefont {Eaves},
  \citenamefont {Loparo}, \citenamefont {Fecko}, \citenamefont {Roberts},
  \citenamefont {Tokmakoff},\ and\ \citenamefont
  {Geissler}}]{a:geissler-tokmakoff-2005}%
  \BibitemOpen
  \bibfield  {author} {\bibinfo {author} {\bibfnamefont {J.~D.}\ \bibnamefont
  {Eaves}}, \bibinfo {author} {\bibfnamefont {J.~J.}\ \bibnamefont {Loparo}},
  \bibinfo {author} {\bibfnamefont {C.~J.}\ \bibnamefont {Fecko}}, \bibinfo
  {author} {\bibfnamefont {S.~T.}\ \bibnamefont {Roberts}}, \bibinfo {author}
  {\bibfnamefont {A.}~\bibnamefont {Tokmakoff}}, \ and\ \bibinfo {author}
  {\bibfnamefont {P.~L.}\ \bibnamefont {Geissler}},\ }\bibfield  {title}
  {\enquote {\bibinfo {title} {Hydrogen bonds in liquid water are broken only
  fleetingly},}\ }\href@noop {} {\bibfield  {journal} {\bibinfo  {journal}
  {Proc. Natl. Acad. Sci. U. S. A.},\ }\textbf {\bibinfo {volume} {102}},\
  \bibinfo {pages} {13019--13022} (\bibinfo {year} {2005})}\BibitemShut
  {NoStop}%
\bibitem [{\citenamefont {Smith}\ \emph
  {et~al.}(2005){\natexlab{a}}\citenamefont {Smith}, \citenamefont {Cappa},
  \citenamefont {Wilson}, \citenamefont {Cohen}, \citenamefont {Geissler},\
  and\ \citenamefont {Saykally}}]{a:unified-saykally-2005}%
  \BibitemOpen
  \bibfield  {author} {\bibinfo {author} {\bibfnamefont {J.~D.}\ \bibnamefont
  {Smith}}, \bibinfo {author} {\bibfnamefont {C.~D.}\ \bibnamefont {Cappa}},
  \bibinfo {author} {\bibfnamefont {K.~R.}\ \bibnamefont {Wilson}}, \bibinfo
  {author} {\bibfnamefont {R.~C.}\ \bibnamefont {Cohen}}, \bibinfo {author}
  {\bibfnamefont {P.~L.}\ \bibnamefont {Geissler}}, \ and\ \bibinfo {author}
  {\bibfnamefont {R.~J.}\ \bibnamefont {Saykally}},\ }\bibfield  {title}
  {\enquote {\bibinfo {title} {Unified description of temperature-dependent
  hydrogen-bond rearrangements in liquid water},}\ }\href@noop {} {\bibfield
  {journal} {\bibinfo  {journal} {Proc. Natl. Acad. Sci. U. S. A.},\ }\textbf
  {\bibinfo {volume} {102}},\ \bibinfo {pages} {14171--14174} (\bibinfo {year}
  {2005}{\natexlab{a}})}\BibitemShut {NoStop}%
\bibitem [{\citenamefont {Bakker}\ and\ \citenamefont
  {Skinner}(2010)}]{a:skinner-review-2010}%
  \BibitemOpen
  \bibfield  {author} {\bibinfo {author} {\bibfnamefont {H.~J.}\ \bibnamefont
  {Bakker}}\ and\ \bibinfo {author} {\bibfnamefont {J.~L.}\ \bibnamefont
  {Skinner}},\ }\bibfield  {title} {\enquote {\bibinfo {title} {Vibrational
  spectroscopy as a probe of structure and dynamics in liquid water},}\
  }\href@noop {} {\bibfield  {journal} {\bibinfo  {journal} {Chem. Rev.},\
  }\textbf {\bibinfo {volume} {110}},\ \bibinfo {pages} {1498--1517} (\bibinfo
  {year} {2010})}\BibitemShut {NoStop}%
\bibitem [{\citenamefont {Eisenberg}\ and\ \citenamefont
  {Kauzmann}(1969)}]{b:water1969}%
  \BibitemOpen
  \bibfield  {author} {\bibinfo {author} {\bibfnamefont {D.}~\bibnamefont
  {Eisenberg}}\ and\ \bibinfo {author} {\bibfnamefont {W.}~\bibnamefont
  {Kauzmann}},\ }\href@noop {} {\emph {\bibinfo {title} {The Structure and
  Properties of Water}}}\ (\bibinfo  {publisher} {Clarendon},\ \bibinfo
  {address} {Oxford},\ \bibinfo {year} {1969})\BibitemShut {NoStop}%
\bibitem [{\citenamefont {Todorova}\ \emph {et~al.}(2006)\citenamefont
  {Todorova}, \citenamefont {Seitsonen}, \citenamefont {Hutter}, \citenamefont
  {Kuo},\ and\ \citenamefont {Mundy}}]{a:hutter-water-hybrid-dft}%
  \BibitemOpen
  \bibfield  {author} {\bibinfo {author} {\bibfnamefont {T.}~\bibnamefont
  {Todorova}}, \bibinfo {author} {\bibfnamefont {A.~P.}\ \bibnamefont
  {Seitsonen}}, \bibinfo {author} {\bibfnamefont {J.}~\bibnamefont {Hutter}},
  \bibinfo {author} {\bibfnamefont {I.~F.~W.}\ \bibnamefont {Kuo}}, \ and\
  \bibinfo {author} {\bibfnamefont {C.~J.}\ \bibnamefont {Mundy}},\ }\bibfield
  {title} {\enquote {\bibinfo {title} {Molecular dynamics simulation of liquid
  water: Hybrid density functionals},}\ }\href@noop {} {\bibfield  {journal}
  {\bibinfo  {journal} {J. Phys. Chem. B},\ }\textbf {\bibinfo {volume}
  {110}},\ \bibinfo {pages} {3685--3691} (\bibinfo {year} {2006})}\BibitemShut
  {NoStop}%
\bibitem [{\citenamefont {Lee}\ and\ \citenamefont
  {Tuckerman}(2006)}]{a:tuckerman-water-2006}%
  \BibitemOpen
  \bibfield  {author} {\bibinfo {author} {\bibfnamefont {H.~S.}\ \bibnamefont
  {Lee}}\ and\ \bibinfo {author} {\bibfnamefont {M.~E.}\ \bibnamefont
  {Tuckerman}},\ }\bibfield  {title} {\enquote {\bibinfo {title} {Structure of
  liquid water at ambient temperature from ab initio molecular dynamics
  performed in the complete basis set limit},}\ }\href@noop {} {\bibfield
  {journal} {\bibinfo  {journal} {J. Chem. Phys.},\ }\textbf {\bibinfo {volume}
  {125}},\ \bibinfo {pages} {154507} (\bibinfo {year} {2006})}\BibitemShut
  {NoStop}%
\bibitem [{\citenamefont {Paesani}\ and\ \citenamefont
  {Voth}(2009)}]{a:voth-water-review-2009}%
  \BibitemOpen
  \bibfield  {author} {\bibinfo {author} {\bibfnamefont {F.}~\bibnamefont
  {Paesani}}\ and\ \bibinfo {author} {\bibfnamefont {G.~A.}\ \bibnamefont
  {Voth}},\ }\bibfield  {title} {\enquote {\bibinfo {title} {The properties of
  water: Insights from quantum simulations},}\ }\href@noop {} {\bibfield
  {journal} {\bibinfo  {journal} {J. Phys. Chem. B},\ }\textbf {\bibinfo
  {volume} {113}},\ \bibinfo {pages} {5702--5719} (\bibinfo {year}
  {2009})}\BibitemShut {NoStop}%
\bibitem [{\citenamefont {K{\"u}hne}\ \emph {et~al.}(2009)\citenamefont
  {K{\"u}hne}, \citenamefont {Krack},\ and\ \citenamefont
  {Parrinello}}]{a:kuhnewater}%
  \BibitemOpen
  \bibfield  {author} {\bibinfo {author} {\bibfnamefont {T.~D.}\ \bibnamefont
  {K{\"u}hne}}, \bibinfo {author} {\bibfnamefont {M.}~\bibnamefont {Krack}}, \
  and\ \bibinfo {author} {\bibfnamefont {M.}~\bibnamefont {Parrinello}},\
  }\bibfield  {title} {\enquote {\bibinfo {title} {Static and dynamical
  properties of liquid water from first principles by a novel
  {C}ar-{P}arrinello-like approach},}\ }\href@noop {} {\bibfield  {journal}
  {\bibinfo  {journal} {J. Chem. Theory Comput.},\ }\textbf {\bibinfo {volume}
  {5}},\ \bibinfo {pages} {235--241} (\bibinfo {year} {2009})}\BibitemShut
  {NoStop}%
\bibitem [{\citenamefont {Wernet}\ \emph {et~al.}(2004)\citenamefont {Wernet},
  \citenamefont {Nordlund}, \citenamefont {Bergmann}, \citenamefont
  {Cavalleri}, \citenamefont {Odelius}, \citenamefont {Ogasawara},
  \citenamefont {Naslund}, \citenamefont {Hirsch}, \citenamefont {Ojamae},
  \citenamefont {Glatzel}, \citenamefont {Pettersson},\ and\ \citenamefont
  {Nilsson}}]{a:nilsson}%
  \BibitemOpen
  \bibfield  {author} {\bibinfo {author} {\bibfnamefont {P.}~\bibnamefont
  {Wernet}}, \bibinfo {author} {\bibfnamefont {D.}~\bibnamefont {Nordlund}},
  \bibinfo {author} {\bibfnamefont {U.}~\bibnamefont {Bergmann}}, \bibinfo
  {author} {\bibfnamefont {M.}~\bibnamefont {Cavalleri}}, \bibinfo {author}
  {\bibfnamefont {M.}~\bibnamefont {Odelius}}, \bibinfo {author} {\bibfnamefont
  {H.}~\bibnamefont {Ogasawara}}, \bibinfo {author} {\bibfnamefont {L.~A.}\
  \bibnamefont {Naslund}}, \bibinfo {author} {\bibfnamefont {T.~K.}\
  \bibnamefont {Hirsch}}, \bibinfo {author} {\bibfnamefont {L.}~\bibnamefont
  {Ojamae}}, \bibinfo {author} {\bibfnamefont {P.}~\bibnamefont {Glatzel}},
  \bibinfo {author} {\bibfnamefont {L.~G.~M.}\ \bibnamefont {Pettersson}}, \
  and\ \bibinfo {author} {\bibfnamefont {A.}~\bibnamefont {Nilsson}},\
  }\bibfield  {title} {\enquote {\bibinfo {title} {The structure of the first
  coordination shell in liquid water},}\ }\href@noop {} {\bibfield  {journal}
  {\bibinfo  {journal} {Science},\ }\textbf {\bibinfo {volume} {304}},\
  \bibinfo {pages} {995--999} (\bibinfo {year} {2004})}\BibitemShut {NoStop}%
\bibitem [{\citenamefont {Tokushima}\ \emph {et~al.}(2008)\citenamefont
  {Tokushima}, \citenamefont {Harada}, \citenamefont {Takahashi}, \citenamefont
  {Senba}, \citenamefont {Ohashi}, \citenamefont {Pettersson}, \citenamefont
  {Nilsson},\ and\ \citenamefont {Shin}}]{a:water-xes-1}%
  \BibitemOpen
  \bibfield  {author} {\bibinfo {author} {\bibfnamefont {T.}~\bibnamefont
  {Tokushima}}, \bibinfo {author} {\bibfnamefont {Y.}~\bibnamefont {Harada}},
  \bibinfo {author} {\bibfnamefont {O.}~\bibnamefont {Takahashi}}, \bibinfo
  {author} {\bibfnamefont {Y.}~\bibnamefont {Senba}}, \bibinfo {author}
  {\bibfnamefont {H.}~\bibnamefont {Ohashi}}, \bibinfo {author} {\bibfnamefont
  {L.~G.~M.}\ \bibnamefont {Pettersson}}, \bibinfo {author} {\bibfnamefont
  {A.}~\bibnamefont {Nilsson}}, \ and\ \bibinfo {author} {\bibfnamefont
  {S.}~\bibnamefont {Shin}},\ }\bibfield  {title} {\enquote {\bibinfo {title}
  {High resolution {X}-ray emission spectroscopy of liquid water: The
  observation of two structural motifs},}\ }\href@noop {} {\bibfield  {journal}
  {\bibinfo  {journal} {Chem. Phys. Lett.},\ }\textbf {\bibinfo {volume}
  {460}},\ \bibinfo {pages} {387--400} (\bibinfo {year} {2008})}\BibitemShut
  {NoStop}%
\bibitem [{\citenamefont {Tokushima}\ \emph {et~al.}(2010)\citenamefont
  {Tokushima}, \citenamefont {Harada}, \citenamefont {Horikawa}, \citenamefont
  {Takahashi}, \citenamefont {Senba}, \citenamefont {Ohashi}, \citenamefont
  {Pettersson}, \citenamefont {Nilsson},\ and\ \citenamefont
  {Shin}}]{a:water-xes-2}%
  \BibitemOpen
  \bibfield  {author} {\bibinfo {author} {\bibfnamefont {T.}~\bibnamefont
  {Tokushima}}, \bibinfo {author} {\bibfnamefont {Y.}~\bibnamefont {Harada}},
  \bibinfo {author} {\bibfnamefont {Y.}~\bibnamefont {Horikawa}}, \bibinfo
  {author} {\bibfnamefont {O.}~\bibnamefont {Takahashi}}, \bibinfo {author}
  {\bibfnamefont {Y.}~\bibnamefont {Senba}}, \bibinfo {author} {\bibfnamefont
  {H.}~\bibnamefont {Ohashi}}, \bibinfo {author} {\bibfnamefont {L.~G.~M.}\
  \bibnamefont {Pettersson}}, \bibinfo {author} {\bibfnamefont
  {A.}~\bibnamefont {Nilsson}}, \ and\ \bibinfo {author} {\bibfnamefont
  {S.}~\bibnamefont {Shin}},\ }\bibfield  {title} {\enquote {\bibinfo {title}
  {High resolution {X}-ray emission spectroscopy of water and its assignment
  based on two structural motifs},}\ }\href@noop {} {\bibfield  {journal}
  {\bibinfo  {journal} {J. Electron Spectrosc. Relat. Phenom.},\ }\textbf
  {\bibinfo {volume} {177}},\ \bibinfo {pages} {192--205} (\bibinfo {year}
  {2010})}\BibitemShut {NoStop}%
\bibitem [{\citenamefont {Tokushima}\ \emph {et~al.}(2012)\citenamefont
  {Tokushima}, \citenamefont {Horikawa}, \citenamefont {Arai}, \citenamefont
  {Harada}, \citenamefont {Takahashi}, \citenamefont {Pettersson},
  \citenamefont {Nilsson},\ and\ \citenamefont {Shin}}]{a:water-xes-3}%
  \BibitemOpen
  \bibfield  {author} {\bibinfo {author} {\bibfnamefont {T.}~\bibnamefont
  {Tokushima}}, \bibinfo {author} {\bibfnamefont {Y.}~\bibnamefont {Horikawa}},
  \bibinfo {author} {\bibfnamefont {H.}~\bibnamefont {Arai}}, \bibinfo {author}
  {\bibfnamefont {Y.}~\bibnamefont {Harada}}, \bibinfo {author} {\bibfnamefont
  {O.}~\bibnamefont {Takahashi}}, \bibinfo {author} {\bibfnamefont {L.~G.~M.}\
  \bibnamefont {Pettersson}}, \bibinfo {author} {\bibfnamefont
  {A.}~\bibnamefont {Nilsson}}, \ and\ \bibinfo {author} {\bibfnamefont
  {S.}~\bibnamefont {Shin}},\ }\bibfield  {title} {\enquote {\bibinfo {title}
  {Polarization dependent resonant x-ray emission spectroscopy of {D2O} and
  {H2O} water: Assignment of the local molecular orbital symmetry},}\
  }\href@noop {} {\bibfield  {journal} {\bibinfo  {journal} {J. Chem. Phys.},\
  }\textbf {\bibinfo {volume} {136}},\ \bibinfo {pages} {044517} (\bibinfo
  {year} {2012})}\BibitemShut {NoStop}%
\bibitem [{\citenamefont {Odelius}\ \emph {et~al.}(2006)\citenamefont
  {Odelius}, \citenamefont {Cavalleri}, \citenamefont {Nilsson},\ and\
  \citenamefont {Pettersson}}]{a:simulation-xray-nilsson}%
  \BibitemOpen
  \bibfield  {author} {\bibinfo {author} {\bibfnamefont {M.}~\bibnamefont
  {Odelius}}, \bibinfo {author} {\bibfnamefont {M.}~\bibnamefont {Cavalleri}},
  \bibinfo {author} {\bibfnamefont {A.}~\bibnamefont {Nilsson}}, \ and\
  \bibinfo {author} {\bibfnamefont {L.~G.~M.}\ \bibnamefont {Pettersson}},\
  }\bibfield  {title} {\enquote {\bibinfo {title} {X-ray absorption spectrum of
  liquid water from molecular dynamics simulations: Asymmetric model},}\
  }\href@noop {} {\bibfield  {journal} {\bibinfo  {journal} {Phys. Rev. B},\
  }\textbf {\bibinfo {volume} {73}},\ \bibinfo {pages} {024205} (\bibinfo
  {year} {2006})}\BibitemShut {NoStop}%
\bibitem [{\citenamefont {Nilsson}\ and\ \citenamefont
  {Pettersson}(2011)}]{a:nilsson-pettersson-perspective}%
  \BibitemOpen
  \bibfield  {author} {\bibinfo {author} {\bibfnamefont {A.}~\bibnamefont
  {Nilsson}}\ and\ \bibinfo {author} {\bibfnamefont {L.~G.~M.}\ \bibnamefont
  {Pettersson}},\ }\bibfield  {title} {\enquote {\bibinfo {title} {Perspective
  on the structure of liquid water},}\ }\href@noop {} {\bibfield  {journal}
  {\bibinfo  {journal} {Chem. Phys.},\ }\textbf {\bibinfo {volume} {389}},\
  \bibinfo {pages} {1--34} (\bibinfo {year} {2011})}\BibitemShut {NoStop}%
\bibitem [{\citenamefont {Huang}\ \emph {et~al.}(2009)\citenamefont {Huang},
  \citenamefont {Wikfeldt}, \citenamefont {Tokushima}, \citenamefont
  {Nordlund}, \citenamefont {Harada}, \citenamefont {Bergmann}, \citenamefont
  {Niebuhr}, \citenamefont {Weiss}, \citenamefont {Horikawa}, \citenamefont
  {Leetmaa}, \citenamefont {Ljungberg}, \citenamefont {Takahashi},
  \citenamefont {Lenz}, \citenamefont {Ojamae}, \citenamefont {Lyubartsev},
  \citenamefont {Shin}, \citenamefont {Pettersson},\ and\ \citenamefont
  {Nilsson}}]{a:water-saxs-1}%
  \BibitemOpen
  \bibfield  {author} {\bibinfo {author} {\bibfnamefont {C.}~\bibnamefont
  {Huang}}, \bibinfo {author} {\bibfnamefont {K.~T.}\ \bibnamefont {Wikfeldt}},
  \bibinfo {author} {\bibfnamefont {T.}~\bibnamefont {Tokushima}}, \bibinfo
  {author} {\bibfnamefont {D.}~\bibnamefont {Nordlund}}, \bibinfo {author}
  {\bibfnamefont {Y.}~\bibnamefont {Harada}}, \bibinfo {author} {\bibfnamefont
  {U.}~\bibnamefont {Bergmann}}, \bibinfo {author} {\bibfnamefont
  {M.}~\bibnamefont {Niebuhr}}, \bibinfo {author} {\bibfnamefont {T.~M.}\
  \bibnamefont {Weiss}}, \bibinfo {author} {\bibfnamefont {Y.}~\bibnamefont
  {Horikawa}}, \bibinfo {author} {\bibfnamefont {M.}~\bibnamefont {Leetmaa}},
  \bibinfo {author} {\bibfnamefont {M.~P.}\ \bibnamefont {Ljungberg}}, \bibinfo
  {author} {\bibfnamefont {O.}~\bibnamefont {Takahashi}}, \bibinfo {author}
  {\bibfnamefont {A.}~\bibnamefont {Lenz}}, \bibinfo {author} {\bibfnamefont
  {L.}~\bibnamefont {Ojamae}}, \bibinfo {author} {\bibfnamefont {A.~P.}\
  \bibnamefont {Lyubartsev}}, \bibinfo {author} {\bibfnamefont
  {S.}~\bibnamefont {Shin}}, \bibinfo {author} {\bibfnamefont {L.~G.~M.}\
  \bibnamefont {Pettersson}}, \ and\ \bibinfo {author} {\bibfnamefont
  {A.}~\bibnamefont {Nilsson}},\ }\bibfield  {title} {\enquote {\bibinfo
  {title} {The inhomogeneous structure of water at ambient conditions},}\
  }\href@noop {} {\bibfield  {journal} {\bibinfo  {journal} {Proc. Natl. Acad.
  Sci. U. S. A.},\ }\textbf {\bibinfo {volume} {106}},\ \bibinfo {pages}
  {15214--15218} (\bibinfo {year} {2009})}\BibitemShut {NoStop}%
\bibitem [{\citenamefont {Smith}\ \emph {et~al.}(2004)\citenamefont {Smith},
  \citenamefont {Cappa}, \citenamefont {Wilson}, \citenamefont {Messer},
  \citenamefont {Cohen},\ and\ \citenamefont {Saykally}}]{a:saykally0}%
  \BibitemOpen
  \bibfield  {author} {\bibinfo {author} {\bibfnamefont {J.~D.}\ \bibnamefont
  {Smith}}, \bibinfo {author} {\bibfnamefont {C.~D.}\ \bibnamefont {Cappa}},
  \bibinfo {author} {\bibfnamefont {K.~R.}\ \bibnamefont {Wilson}}, \bibinfo
  {author} {\bibfnamefont {B.~M.}\ \bibnamefont {Messer}}, \bibinfo {author}
  {\bibfnamefont {R.~C.}\ \bibnamefont {Cohen}}, \ and\ \bibinfo {author}
  {\bibfnamefont {R.~J.}\ \bibnamefont {Saykally}},\ }\bibfield  {title}
  {\enquote {\bibinfo {title} {Energetics of hydrogen bond network
  rearrangements in liquid water},}\ }\href@noop {} {\bibfield  {journal}
  {\bibinfo  {journal} {Science},\ }\textbf {\bibinfo {volume} {306}},\
  \bibinfo {pages} {851--853} (\bibinfo {year} {2004})}\BibitemShut {NoStop}%
\bibitem [{\citenamefont {Smith}\ \emph {et~al.}(2006)\citenamefont {Smith},
  \citenamefont {Cappa}, \citenamefont {Messer}, \citenamefont {Drisdell},
  \citenamefont {Cohen},\ and\ \citenamefont {Saykally}}]{a:saykally1}%
  \BibitemOpen
  \bibfield  {author} {\bibinfo {author} {\bibfnamefont {J.~D.}\ \bibnamefont
  {Smith}}, \bibinfo {author} {\bibfnamefont {C.~D.}\ \bibnamefont {Cappa}},
  \bibinfo {author} {\bibfnamefont {B.~M.}\ \bibnamefont {Messer}}, \bibinfo
  {author} {\bibfnamefont {W.~S.}\ \bibnamefont {Drisdell}}, \bibinfo {author}
  {\bibfnamefont {R.~C.}\ \bibnamefont {Cohen}}, \ and\ \bibinfo {author}
  {\bibfnamefont {R.~J.}\ \bibnamefont {Saykally}},\ }\bibfield  {title}
  {\enquote {\bibinfo {title} {Probing the local structure of liquid water by
  {X}-ray absorption spectroscopy},}\ }\href@noop {} {\bibfield  {journal}
  {\bibinfo  {journal} {J. Phys. Chem. B},\ }\textbf {\bibinfo {volume}
  {110}},\ \bibinfo {pages} {20038--20045} (\bibinfo {year}
  {2006})}\BibitemShut {NoStop}%
\bibitem [{\citenamefont {Head-Gordon}\ and\ \citenamefont
  {Johnson}(2006)}]{a:thg}%
  \BibitemOpen
  \bibfield  {author} {\bibinfo {author} {\bibfnamefont {T.}~\bibnamefont
  {Head-Gordon}}\ and\ \bibinfo {author} {\bibfnamefont {M.~E.}\ \bibnamefont
  {Johnson}},\ }\bibfield  {title} {\enquote {\bibinfo {title} {Tetrahedral
  structure or chains for liquid water},}\ }\href@noop {} {\bibfield  {journal}
  {\bibinfo  {journal} {Proc. Natl. Acad. Sci. U. S. A.},\ }\textbf {\bibinfo
  {volume} {103}},\ \bibinfo {pages} {7973--7977} (\bibinfo {year}
  {2006})}\BibitemShut {NoStop}%
\bibitem [{\citenamefont {Head-Gordon}\ and\ \citenamefont
  {Rick}(2007)}]{a:thg1}%
  \BibitemOpen
  \bibfield  {author} {\bibinfo {author} {\bibfnamefont {T.}~\bibnamefont
  {Head-Gordon}}\ and\ \bibinfo {author} {\bibfnamefont {S.~W.}\ \bibnamefont
  {Rick}},\ }\bibfield  {title} {\enquote {\bibinfo {title} {Consequences of
  chain networks on thermodynamic, dielectric and structural properties for
  liquid water},}\ }\href@noop {} {\bibfield  {journal} {\bibinfo  {journal}
  {Phys. Chem. Chem. Phys.},\ }\textbf {\bibinfo {volume} {9}},\ \bibinfo
  {pages} {83--91} (\bibinfo {year} {2007})}\BibitemShut {NoStop}%
\bibitem [{\citenamefont {Clark}\ \emph
  {et~al.}(2010){\natexlab{b}}\citenamefont {Clark}, \citenamefont {Hura},
  \citenamefont {Teixeira}, \citenamefont {Soper},\ and\ \citenamefont
  {Head-Gordon}}]{a:water-saxs-thg}%
  \BibitemOpen
  \bibfield  {author} {\bibinfo {author} {\bibfnamefont {G.~N.~I.}\
  \bibnamefont {Clark}}, \bibinfo {author} {\bibfnamefont {G.~L.}\ \bibnamefont
  {Hura}}, \bibinfo {author} {\bibfnamefont {J.}~\bibnamefont {Teixeira}},
  \bibinfo {author} {\bibfnamefont {A.~K.}\ \bibnamefont {Soper}}, \ and\
  \bibinfo {author} {\bibfnamefont {T.}~\bibnamefont {Head-Gordon}},\
  }\bibfield  {title} {\enquote {\bibinfo {title} {Small-angle scattering and
  the structure of ambient liquid water},}\ }\href@noop {} {\bibfield
  {journal} {\bibinfo  {journal} {Proc. Natl. Acad. Sci. U. S. A.},\ }\textbf
  {\bibinfo {volume} {107}},\ \bibinfo {pages} {14003--14007} (\bibinfo {year}
  {2010}{\natexlab{b}})}\BibitemShut {NoStop}%
\bibitem [{\citenamefont {Prendergast}\ and\ \citenamefont
  {Galli}(2006)}]{a:xas-prendergast-galli}%
  \BibitemOpen
  \bibfield  {author} {\bibinfo {author} {\bibfnamefont {D.}~\bibnamefont
  {Prendergast}}\ and\ \bibinfo {author} {\bibfnamefont {G.}~\bibnamefont
  {Galli}},\ }\bibfield  {title} {\enquote {\bibinfo {title} {X-ray absorption
  spectra of water from first principles calculations},}\ }\href@noop {}
  {\bibfield  {journal} {\bibinfo  {journal} {Phys. Rev. Lett.},\ }\textbf
  {\bibinfo {volume} {96}},\ \bibinfo {pages} {215502} (\bibinfo {year}
  {2006})}\BibitemShut {NoStop}%
\bibitem [{\citenamefont {Fernandez-Serra}\ and\ \citenamefont
  {Artacho}(2006)}]{a:water-artacho}%
  \BibitemOpen
  \bibfield  {author} {\bibinfo {author} {\bibfnamefont {M.~V.}\ \bibnamefont
  {Fernandez-Serra}}\ and\ \bibinfo {author} {\bibfnamefont {E.}~\bibnamefont
  {Artacho}},\ }\bibfield  {title} {\enquote {\bibinfo {title} {Electrons and
  hydrogen-bond connectivity in liquid water},}\ }\href@noop {} {\bibfield
  {journal} {\bibinfo  {journal} {Phys. Rev. Lett.},\ }\textbf {\bibinfo
  {volume} {96}},\ \bibinfo {pages} {016404} (\bibinfo {year}
  {2006})}\BibitemShut {NoStop}%
\bibitem [{\citenamefont {Soper}(2010)}]{a:soper-myths}%
  \BibitemOpen
  \bibfield  {author} {\bibinfo {author} {\bibfnamefont {A.~K.}\ \bibnamefont
  {Soper}},\ }\bibfield  {title} {\enquote {\bibinfo {title} {Recent water
  myths},}\ }\href@noop {} {\bibfield  {journal} {\bibinfo  {journal} {Pure
  Appl. Chem.},\ }\textbf {\bibinfo {volume} {82}},\ \bibinfo {pages}
  {1855--1867} (\bibinfo {year} {2010})}\BibitemShut {NoStop}%
\bibitem [{\citenamefont {Chen}\ \emph {et~al.}(2010)\citenamefont {Chen},
  \citenamefont {Wu},\ and\ \citenamefont {Car}}]{a:car-xas-water}%
  \BibitemOpen
  \bibfield  {author} {\bibinfo {author} {\bibfnamefont {W.}~\bibnamefont
  {Chen}}, \bibinfo {author} {\bibfnamefont {X.~F.}\ \bibnamefont {Wu}}, \ and\
  \bibinfo {author} {\bibfnamefont {R.}~\bibnamefont {Car}},\ }\bibfield
  {title} {\enquote {\bibinfo {title} {X-ray absorption signatures of the
  molecular environment in water and ice},}\ }\href@noop {} {\bibfield
  {journal} {\bibinfo  {journal} {Phys. Rev. Lett.},\ }\textbf {\bibinfo
  {volume} {105}},\ \bibinfo {pages} {017802} (\bibinfo {year}
  {2010})}\BibitemShut {NoStop}%
\bibitem [{\citenamefont {Nilsson}\ \emph {et~al.}(2005)\citenamefont
  {Nilsson}, \citenamefont {Wernet}, \citenamefont {Nordlund}, \citenamefont
  {Bergmann}, \citenamefont {Cavalleri}, \citenamefont {Odelius}, \citenamefont
  {Ogasawara}, \citenamefont {Naslund}, \citenamefont {Hirsch}, \citenamefont
  {Ojamae}, \citenamefont {Glatzel},\ and\ \citenamefont
  {Pettersson}}]{a:nilsson-comment1}%
  \BibitemOpen
  \bibfield  {author} {\bibinfo {author} {\bibfnamefont {A.}~\bibnamefont
  {Nilsson}}, \bibinfo {author} {\bibfnamefont {P.}~\bibnamefont {Wernet}},
  \bibinfo {author} {\bibfnamefont {D.}~\bibnamefont {Nordlund}}, \bibinfo
  {author} {\bibfnamefont {U.}~\bibnamefont {Bergmann}}, \bibinfo {author}
  {\bibfnamefont {M.}~\bibnamefont {Cavalleri}}, \bibinfo {author}
  {\bibfnamefont {M.}~\bibnamefont {Odelius}}, \bibinfo {author} {\bibfnamefont
  {H.}~\bibnamefont {Ogasawara}}, \bibinfo {author} {\bibfnamefont {L.~A.}\
  \bibnamefont {Naslund}}, \bibinfo {author} {\bibfnamefont {T.~K.}\
  \bibnamefont {Hirsch}}, \bibinfo {author} {\bibfnamefont {L.}~\bibnamefont
  {Ojamae}}, \bibinfo {author} {\bibfnamefont {P.}~\bibnamefont {Glatzel}}, \
  and\ \bibinfo {author} {\bibfnamefont {L.~G.~M.}\ \bibnamefont
  {Pettersson}},\ }\bibfield  {title} {\enquote {\bibinfo {title} {Comment on
  "energetics of hydrogen bond network: Rearrangements in liquid water"},}\
  }\href@noop {} {\bibfield  {journal} {\bibinfo  {journal} {Science},\
  }\textbf {\bibinfo {volume} {308}},\ \bibinfo {pages} {793A--793A} (\bibinfo
  {year} {2005})}\BibitemShut {NoStop}%
\bibitem [{\citenamefont {Smith}\ \emph
  {et~al.}(2005){\natexlab{b}}\citenamefont {Smith}, \citenamefont {Cappa},
  \citenamefont {Messer}, \citenamefont {Cohen},\ and\ \citenamefont
  {Saykally}}]{a:saykally-response1}%
  \BibitemOpen
  \bibfield  {author} {\bibinfo {author} {\bibfnamefont {J.~D.}\ \bibnamefont
  {Smith}}, \bibinfo {author} {\bibfnamefont {C.~D.}\ \bibnamefont {Cappa}},
  \bibinfo {author} {\bibfnamefont {B.~M.}\ \bibnamefont {Messer}}, \bibinfo
  {author} {\bibfnamefont {R.~C.}\ \bibnamefont {Cohen}}, \ and\ \bibinfo
  {author} {\bibfnamefont {R.~J.}\ \bibnamefont {Saykally}},\ }\bibfield
  {title} {\enquote {\bibinfo {title} {Response to comment on "energetics of
  hydrogen bond network: Rearrangements in liquid water"},}\ }\href@noop {}
  {\bibfield  {journal} {\bibinfo  {journal} {Science},\ }\textbf {\bibinfo
  {volume} {308}},\ \bibinfo {pages} {793B--793B} (\bibinfo {year}
  {2005}{\natexlab{b}})}\BibitemShut {NoStop}%
\bibitem [{\citenamefont {Soper}(2005)}]{a:soper}%
  \BibitemOpen
  \bibfield  {author} {\bibinfo {author} {\bibfnamefont {A.~K.}\ \bibnamefont
  {Soper}},\ }\bibfield  {title} {\enquote {\bibinfo {title} {An asymmetric
  model for water structure},}\ }\href@noop {} {\bibfield  {journal} {\bibinfo
  {journal} {J. Phys.: Condens. Matter},\ }\textbf {\bibinfo {volume} {17}},\
  \bibinfo {pages} {S3273--S3282} (\bibinfo {year} {2005})}\BibitemShut
  {NoStop}%
\bibitem [{\citenamefont {Leetmaa}\ \emph {et~al.}(2008)\citenamefont
  {Leetmaa}, \citenamefont {Wikfeldt}, \citenamefont {Ljungberg}, \citenamefont
  {Odelius}, \citenamefont {Swenson}, \citenamefont {Nilsson},\ and\
  \citenamefont {Pettersson}}]{a:water-chains-rmc}%
  \BibitemOpen
  \bibfield  {author} {\bibinfo {author} {\bibfnamefont {M.}~\bibnamefont
  {Leetmaa}}, \bibinfo {author} {\bibfnamefont {K.~T.}\ \bibnamefont
  {Wikfeldt}}, \bibinfo {author} {\bibfnamefont {M.~P.}\ \bibnamefont
  {Ljungberg}}, \bibinfo {author} {\bibfnamefont {M.}~\bibnamefont {Odelius}},
  \bibinfo {author} {\bibfnamefont {J.}~\bibnamefont {Swenson}}, \bibinfo
  {author} {\bibfnamefont {A.}~\bibnamefont {Nilsson}}, \ and\ \bibinfo
  {author} {\bibfnamefont {L.~G.~M.}\ \bibnamefont {Pettersson}},\ }\bibfield
  {title} {\enquote {\bibinfo {title} {Diffraction and {IR/Raman} data do not
  prove tetrahedral water},}\ }\href@noop {} {\bibfield  {journal} {\bibinfo
  {journal} {J. Chem. Phys.},\ }\textbf {\bibinfo {volume} {129}},\ \bibinfo
  {pages} {084502} (\bibinfo {year} {2008})}\BibitemShut {NoStop}%
\bibitem [{\citenamefont {Leetmaa}\ \emph {et~al.}(2006)\citenamefont
  {Leetmaa}, \citenamefont {Ljungberg}, \citenamefont {Ogasawara},
  \citenamefont {Odelius}, \citenamefont {Naslund}, \citenamefont {Nilsson},\
  and\ \citenamefont {Pettersson}}]{a:one-more-nilsson}%
  \BibitemOpen
  \bibfield  {author} {\bibinfo {author} {\bibfnamefont {M.}~\bibnamefont
  {Leetmaa}}, \bibinfo {author} {\bibfnamefont {M.}~\bibnamefont {Ljungberg}},
  \bibinfo {author} {\bibfnamefont {H.}~\bibnamefont {Ogasawara}}, \bibinfo
  {author} {\bibfnamefont {M.}~\bibnamefont {Odelius}}, \bibinfo {author}
  {\bibfnamefont {L.~A.}\ \bibnamefont {Naslund}}, \bibinfo {author}
  {\bibfnamefont {A.}~\bibnamefont {Nilsson}}, \ and\ \bibinfo {author}
  {\bibfnamefont {L.~G.~M.}\ \bibnamefont {Pettersson}},\ }\bibfield  {title}
  {\enquote {\bibinfo {title} {Are recent water models obtained by fitting
  diffraction data consistent with infrared/{R}aman and {X}-ray absorption
  spectra?}}\ }\href@noop {} {\bibfield  {journal} {\bibinfo  {journal} {J.
  Chem. Phys.},\ }\textbf {\bibinfo {volume} {125}},\ \bibinfo {pages} {244510}
  (\bibinfo {year} {2006})}\BibitemShut {NoStop}%
\bibitem [{\citenamefont {Wikfeldt}\ \emph {et~al.}(2009)\citenamefont
  {Wikfeldt}, \citenamefont {Leetmaa}, \citenamefont {Ljungberg}, \citenamefont
  {Nilsson},\ and\ \citenamefont {Pettersson}}]{a:water-models-2009}%
  \BibitemOpen
  \bibfield  {author} {\bibinfo {author} {\bibfnamefont {K.~T.}\ \bibnamefont
  {Wikfeldt}}, \bibinfo {author} {\bibfnamefont {M.}~\bibnamefont {Leetmaa}},
  \bibinfo {author} {\bibfnamefont {M.~P.}\ \bibnamefont {Ljungberg}}, \bibinfo
  {author} {\bibfnamefont {A.}~\bibnamefont {Nilsson}}, \ and\ \bibinfo
  {author} {\bibfnamefont {L.~G.~M.}\ \bibnamefont {Pettersson}},\ }\bibfield
  {title} {\enquote {\bibinfo {title} {On the range of water structure models
  compatible with {X}-ray and neutron diffraction data},}\ }\href@noop {}
  {\bibfield  {journal} {\bibinfo  {journal} {J. Phys. Chem. B},\ }\textbf
  {\bibinfo {volume} {113}},\ \bibinfo {pages} {6246--6255} (\bibinfo {year}
  {2009})}\BibitemShut {NoStop}%
\bibitem [{\citenamefont {Khaliullin}\ \emph {et~al.}(2007)\citenamefont
  {Khaliullin}, \citenamefont {Cobar}, \citenamefont {Lochan}, \citenamefont
  {Bell},\ and\ \citenamefont {Head-Gordon}}]{a:theeda}%
  \BibitemOpen
  \bibfield  {author} {\bibinfo {author} {\bibfnamefont {R.~Z.}\ \bibnamefont
  {Khaliullin}}, \bibinfo {author} {\bibfnamefont {E.~A.}\ \bibnamefont
  {Cobar}}, \bibinfo {author} {\bibfnamefont {R.~C.}\ \bibnamefont {Lochan}},
  \bibinfo {author} {\bibfnamefont {A.~T.}\ \bibnamefont {Bell}}, \ and\
  \bibinfo {author} {\bibfnamefont {M.}~\bibnamefont {Head-Gordon}},\
  }\bibfield  {title} {\enquote {\bibinfo {title} {Unravelling the origin of
  intermolecular interactions using absolutely localized molecular orbitals},}\
  }\href@noop {} {\bibfield  {journal} {\bibinfo  {journal} {J. Phys. Chem.
  A},\ }\textbf {\bibinfo {volume} {111}},\ \bibinfo {pages} {8753--8765}
  (\bibinfo {year} {2007})}\BibitemShut {NoStop}%
\bibitem [{\citenamefont {Kitaura}\ and\ \citenamefont
  {Morokuma}(1976)}]{a:km}%
  \BibitemOpen
  \bibfield  {author} {\bibinfo {author} {\bibfnamefont {K.}~\bibnamefont
  {Kitaura}}\ and\ \bibinfo {author} {\bibfnamefont {K.}~\bibnamefont
  {Morokuma}},\ }\bibfield  {title} {\enquote {\bibinfo {title} {New energy
  decomposition scheme for molecular-interactions within {H}artree-{F}ock
  approximation},}\ }\href@noop {} {\bibfield  {journal} {\bibinfo  {journal}
  {Int. J. Quantum Chem.},\ }\textbf {\bibinfo {volume} {10}},\ \bibinfo
  {pages} {325--340} (\bibinfo {year} {1976})}\BibitemShut {NoStop}%
\bibitem [{\citenamefont {Chen}\ and\ \citenamefont {Gordon}(1996)}]{a:rvsX}%
  \BibitemOpen
  \bibfield  {author} {\bibinfo {author} {\bibfnamefont {W.}~\bibnamefont
  {Chen}}\ and\ \bibinfo {author} {\bibfnamefont {M.~S.}\ \bibnamefont
  {Gordon}},\ }\bibfield  {title} {\enquote {\bibinfo {title} {Energy
  decomposition analyses for many-body interaction and applications to water
  complexes},}\ }\href@noop {} {\bibfield  {journal} {\bibinfo  {journal} {J.
  Phys. Chem.},\ }\textbf {\bibinfo {volume} {100}},\ \bibinfo {pages}
  {14316--14328} (\bibinfo {year} {1996})}\BibitemShut {NoStop}%
\bibitem [{\citenamefont {Bagus}\ and\ \citenamefont {Illas}(1992)}]{a:csov1}%
  \BibitemOpen
  \bibfield  {author} {\bibinfo {author} {\bibfnamefont {P.~S.}\ \bibnamefont
  {Bagus}}\ and\ \bibinfo {author} {\bibfnamefont {F.}~\bibnamefont {Illas}},\
  }\bibfield  {title} {\enquote {\bibinfo {title} {Decomposition of the
  chemisorption bond by constrained variations - order of the variations and
  construction of the variational spaces},}\ }\href@noop {} {\bibfield
  {journal} {\bibinfo  {journal} {J. Chem. Phys.},\ }\textbf {\bibinfo {volume}
  {96}},\ \bibinfo {pages} {8962--8970} (\bibinfo {year} {1992})}\BibitemShut
  {NoStop}%
\bibitem [{\citenamefont {Glendening}(2005)}]{a:nedaDFT}%
  \BibitemOpen
  \bibfield  {author} {\bibinfo {author} {\bibfnamefont {E.~D.}\ \bibnamefont
  {Glendening}},\ }\bibfield  {title} {\enquote {\bibinfo {title} {Natural
  energy decomposition analysis: Extension to density functional methods and
  analysis of cooperative effects in water clusters},}\ }\href@noop {}
  {\bibfield  {journal} {\bibinfo  {journal} {J. Phys. Chem. A},\ }\textbf
  {\bibinfo {volume} {109}},\ \bibinfo {pages} {11936--11940} (\bibinfo {year}
  {2005})}\BibitemShut {NoStop}%
\bibitem [{\citenamefont {Mo}\ \emph {et~al.}(2000)\citenamefont {Mo},
  \citenamefont {Gao},\ and\ \citenamefont {Peyerimhoff}}]{a:blweda}%
  \BibitemOpen
  \bibfield  {author} {\bibinfo {author} {\bibfnamefont {Y.~R.}\ \bibnamefont
  {Mo}}, \bibinfo {author} {\bibfnamefont {J.~L.}\ \bibnamefont {Gao}}, \ and\
  \bibinfo {author} {\bibfnamefont {S.~D.}\ \bibnamefont {Peyerimhoff}},\
  }\bibfield  {title} {\enquote {\bibinfo {title} {Energy decomposition
  analysis of intermolecular interactions using a block-localized wave function
  approach},}\ }\href@noop {} {\bibfield  {journal} {\bibinfo  {journal} {J.
  Chem. Phys.},\ }\textbf {\bibinfo {volume} {112}},\ \bibinfo {pages}
  {5530--5538} (\bibinfo {year} {2000})}\BibitemShut {NoStop}%
\bibitem [{\citenamefont {Khaliullin}\ \emph {et~al.}(2008)\citenamefont
  {Khaliullin}, \citenamefont {Bell},\ and\ \citenamefont
  {Head-Gordon}}]{a:cta}%
  \BibitemOpen
  \bibfield  {author} {\bibinfo {author} {\bibfnamefont {R.~Z.}\ \bibnamefont
  {Khaliullin}}, \bibinfo {author} {\bibfnamefont {A.~T.}\ \bibnamefont
  {Bell}}, \ and\ \bibinfo {author} {\bibfnamefont {M.}~\bibnamefont
  {Head-Gordon}},\ }\bibfield  {title} {\enquote {\bibinfo {title} {Analysis of
  charge transfer effects in molecular complexes based on absolutely localized
  molecular orbitals},}\ }\href@noop {} {\bibfield  {journal} {\bibinfo
  {journal} {J. Chem. Phys.},\ }\textbf {\bibinfo {volume} {128}},\ \bibinfo
  {pages} {184112} (\bibinfo {year} {2008})}\BibitemShut {NoStop}%
\bibitem [{\citenamefont {Khaliullin}\ \emph {et~al.}(2009)\citenamefont
  {Khaliullin}, \citenamefont {Bell},\ and\ \citenamefont
  {Head-Gordon}}]{a:khalh2o}%
  \BibitemOpen
  \bibfield  {author} {\bibinfo {author} {\bibfnamefont {R.~Z.}\ \bibnamefont
  {Khaliullin}}, \bibinfo {author} {\bibfnamefont {A.~T.}\ \bibnamefont
  {Bell}}, \ and\ \bibinfo {author} {\bibfnamefont {M.}~\bibnamefont
  {Head-Gordon}},\ }\bibfield  {title} {\enquote {\bibinfo {title} {Electron
  donation in the water-water hydrogen bond},}\ }\href@noop {} {\bibfield
  {journal} {\bibinfo  {journal} {Chem.--Eur. J.},\ }\textbf {\bibinfo {volume}
  {15}},\ \bibinfo {pages} {851--855} (\bibinfo {year} {2009})}\BibitemShut
  {NoStop}%
\bibitem [{\citenamefont {Ramos-Cordoba}\ \emph {et~al.}(2011)\citenamefont
  {Ramos-Cordoba}, \citenamefont {Lambrecht},\ and\ \citenamefont
  {Head-Gordon}}]{a:app_almo_2}%
  \BibitemOpen
  \bibfield  {author} {\bibinfo {author} {\bibfnamefont {E.}~\bibnamefont
  {Ramos-Cordoba}}, \bibinfo {author} {\bibfnamefont {D.~S.}\ \bibnamefont
  {Lambrecht}}, \ and\ \bibinfo {author} {\bibfnamefont {M.}~\bibnamefont
  {Head-Gordon}},\ }\bibfield  {title} {\enquote {\bibinfo {title}
  {Charge-transfer and the hydrogen bond: Spectroscopic and structural
  implications from electronic structure calculations},}\ }\href@noop {}
  {\bibfield  {journal} {\bibinfo  {journal} {Faraday Discuss.},\ }\textbf
  {\bibinfo {volume} {150}},\ \bibinfo {pages} {345--362} (\bibinfo {year}
  {2011})}\BibitemShut {NoStop}%
\bibitem [{\citenamefont {Wang}\ \emph {et~al.}(2010)\citenamefont {Wang},
  \citenamefont {Jenness}, \citenamefont {Al-Saidi},\ and\ \citenamefont
  {Jordan}}]{a:app_almo_4}%
  \BibitemOpen
  \bibfield  {author} {\bibinfo {author} {\bibfnamefont {F.~F.}\ \bibnamefont
  {Wang}}, \bibinfo {author} {\bibfnamefont {G.}~\bibnamefont {Jenness}},
  \bibinfo {author} {\bibfnamefont {W.~A.}\ \bibnamefont {Al-Saidi}}, \ and\
  \bibinfo {author} {\bibfnamefont {K.~D.}\ \bibnamefont {Jordan}},\ }\bibfield
   {title} {\enquote {\bibinfo {title} {Assessment of the performance of common
  density functional methods for describing the interaction energies of
  {(H2O)6} clusters},}\ }\href@noop {} {\bibfield  {journal} {\bibinfo
  {journal} {J. Chem. Phys.},\ }\textbf {\bibinfo {volume} {132}},\ \bibinfo
  {pages} {134303} (\bibinfo {year} {2010})}\BibitemShut {NoStop}%
\bibitem [{\citenamefont {Luck}(1998)}]{a:water-cooperativity0}%
  \BibitemOpen
  \bibfield  {author} {\bibinfo {author} {\bibfnamefont {W.~A.~P.}\
  \bibnamefont {Luck}},\ }\bibfield  {title} {\enquote {\bibinfo {title} {The
  importance of cooperativity for the properties of liquid water},}\
  }\href@noop {} {\bibfield  {journal} {\bibinfo  {journal} {J. Mol. Struct.},\
  }\textbf {\bibinfo {volume} {448}},\ \bibinfo {pages} {131--142} (\bibinfo
  {year} {1998})}\BibitemShut {NoStop}%
\bibitem [{\citenamefont {Kumar}\ \emph {et~al.}(2007)\citenamefont {Kumar},
  \citenamefont {Schmidt},\ and\ \citenamefont {Skinner}}]{a:kumar}%
  \BibitemOpen
  \bibfield  {author} {\bibinfo {author} {\bibfnamefont {R.}~\bibnamefont
  {Kumar}}, \bibinfo {author} {\bibfnamefont {J.~R.}\ \bibnamefont {Schmidt}},
  \ and\ \bibinfo {author} {\bibfnamefont {J.~L.}\ \bibnamefont {Skinner}},\
  }\bibfield  {title} {\enquote {\bibinfo {title} {Hydrogen bonding definitions
  and dynamics in liquid water},}\ }\href@noop {} {\bibfield  {journal}
  {\bibinfo  {journal} {J. Chem. Phys.},\ }\textbf {\bibinfo {volume} {126}},\
  \bibinfo {pages} {204107} (\bibinfo {year} {2007})}\BibitemShut {NoStop}%
\bibitem [{\citenamefont {K{\"u}hne}\ \emph {et~al.}(2007)\citenamefont
  {K{\"u}hne}, \citenamefont {Krack}, \citenamefont {Mohamed},\ and\
  \citenamefont {Parrinello}}]{a:2ndcpmd}%
  \BibitemOpen
  \bibfield  {author} {\bibinfo {author} {\bibfnamefont {T.~D.}\ \bibnamefont
  {K{\"u}hne}}, \bibinfo {author} {\bibfnamefont {M.}~\bibnamefont {Krack}},
  \bibinfo {author} {\bibfnamefont {F.~R.}\ \bibnamefont {Mohamed}}, \ and\
  \bibinfo {author} {\bibfnamefont {M.}~\bibnamefont {Parrinello}},\ }\bibfield
   {title} {\enquote {\bibinfo {title} {Efficient and accurate
  {C}ar-{P}arrinello-like approach to {B}orn-{O}ppenheimer molecular
  dynamics},}\ }\href@noop {} {\bibfield  {journal} {\bibinfo  {journal} {Phys.
  Rev. Lett.},\ }\textbf {\bibinfo {volume} {98}},\ \bibinfo {pages} {066401}
  (\bibinfo {year} {2007})}\BibitemShut {NoStop}%
\bibitem [{\citenamefont {Luzar}\ and\ \citenamefont
  {Chandler}(1996)}]{a:luzar}%
  \BibitemOpen
  \bibfield  {author} {\bibinfo {author} {\bibfnamefont {A.}~\bibnamefont
  {Luzar}}\ and\ \bibinfo {author} {\bibfnamefont {D.}~\bibnamefont
  {Chandler}},\ }\bibfield  {title} {\enquote {\bibinfo {title} {Hydrogen-bond
  kinetics in liquid water},}\ }\href@noop {} {\bibfield  {journal} {\bibinfo
  {journal} {Nature},\ }\textbf {\bibinfo {volume} {379}},\ \bibinfo {pages}
  {55--57} (\bibinfo {year} {1996})}\BibitemShut {NoStop}%
\bibitem [{\citenamefont {Kuo}\ and\ \citenamefont {Mundy}(2004)}]{a:aimd3}%
  \BibitemOpen
  \bibfield  {author} {\bibinfo {author} {\bibfnamefont {I.~F.~W.}\
  \bibnamefont {Kuo}}\ and\ \bibinfo {author} {\bibfnamefont {C.~J.}\
  \bibnamefont {Mundy}},\ }\bibfield  {title} {\enquote {\bibinfo {title} {An
  ab initio molecular dynamics study of the aqueous liquid-vapor interface},}\
  }\href@noop {} {\bibfield  {journal} {\bibinfo  {journal} {Science},\
  }\textbf {\bibinfo {volume} {303}},\ \bibinfo {pages} {658--660} (\bibinfo
  {year} {2004})}\BibitemShut {NoStop}%
\bibitem [{\citenamefont {Leetmaa}\ \emph {et~al.}(2010)\citenamefont
  {Leetmaa}, \citenamefont {Ljungberg}, \citenamefont {Lyubartsev},
  \citenamefont {Nilsson},\ and\ \citenamefont
  {Pettersson}}]{a:hch-water-review}%
  \BibitemOpen
  \bibfield  {author} {\bibinfo {author} {\bibfnamefont {M.}~\bibnamefont
  {Leetmaa}}, \bibinfo {author} {\bibfnamefont {M.~P.}\ \bibnamefont
  {Ljungberg}}, \bibinfo {author} {\bibfnamefont {A.}~\bibnamefont
  {Lyubartsev}}, \bibinfo {author} {\bibfnamefont {A.}~\bibnamefont {Nilsson}},
  \ and\ \bibinfo {author} {\bibfnamefont {L.~G.~M.}\ \bibnamefont
  {Pettersson}},\ }\bibfield  {title} {\enquote {\bibinfo {title} {Theoretical
  approximations to {X}-ray absorption spectroscopy of liquid water and ice},}\
  }\href@noop {} {\bibfield  {journal} {\bibinfo  {journal} {J. Electron
  Spectrosc. Relat. Phenom.},\ }\textbf {\bibinfo {volume} {177}},\ \bibinfo
  {pages} {135--157} (\bibinfo {year} {2010})}\BibitemShut {NoStop}%
\bibitem [{\citenamefont {Iannuzzi}\ and\ \citenamefont
  {Hutter}(2007)}]{a:xas-hutter}%
  \BibitemOpen
  \bibfield  {author} {\bibinfo {author} {\bibfnamefont {M.}~\bibnamefont
  {Iannuzzi}}\ and\ \bibinfo {author} {\bibfnamefont {J.}~\bibnamefont
  {Hutter}},\ }\bibfield  {title} {\enquote {\bibinfo {title} {Inner-shell
  spectroscopy by the gaussian and augmented plane wave method},}\ }\href@noop
  {} {\bibfield  {journal} {\bibinfo  {journal} {Phys. Chem. Chem. Phys.},\
  }\textbf {\bibinfo {volume} {9}},\ \bibinfo {pages} {1599--1610} (\bibinfo
  {year} {2007})}\BibitemShut {NoStop}%
\bibitem [{\citenamefont {Iannuzzi}(2008)}]{a:xas-iannuzzi}%
  \BibitemOpen
  \bibfield  {author} {\bibinfo {author} {\bibfnamefont {M.}~\bibnamefont
  {Iannuzzi}},\ }\bibfield  {title} {\enquote {\bibinfo {title} {X-ray
  absorption spectra of hexagonal ice and liquid water by all-electron gaussian
  and augmented plane wave calculations},}\ }\href@noop {} {\bibfield
  {journal} {\bibinfo  {journal} {J. Chem. Phys.},\ }\textbf {\bibinfo {volume}
  {128}},\ \bibinfo {pages} {204506} (\bibinfo {year} {2008})}\BibitemShut
  {NoStop}%
\bibitem [{\citenamefont {Kong}\ \emph {et~al.}(2012)\citenamefont {Kong},
  \citenamefont {Wu},\ and\ \citenamefont {Car}}]{a:kong-xas}%
  \BibitemOpen
  \bibfield  {author} {\bibinfo {author} {\bibfnamefont {L.}~\bibnamefont
  {Kong}}, \bibinfo {author} {\bibfnamefont {X.}~\bibnamefont {Wu}}, \ and\
  \bibinfo {author} {\bibfnamefont {R.}~\bibnamefont {Car}},\ }\bibfield
  {title} {\enquote {\bibinfo {title} {Roles of quantum nuclei and
  inhomogeneous screening in the {X}-ray absorption spectra of water and
  ice},}\ }\href@noop {} {\bibfield  {journal} {\bibinfo  {journal} {Phys. Rev.
  B},\ }\textbf {\bibinfo {volume} {86}},\ \bibinfo {pages} {134203} (\bibinfo
  {year} {2012})}\BibitemShut {NoStop}%
\bibitem [{\citenamefont {Khaliullin}\ \emph {et~al.}(2006)\citenamefont
  {Khaliullin}, \citenamefont {Head-Gordon},\ and\ \citenamefont
  {Bell}}]{a:khal}%
  \BibitemOpen
  \bibfield  {author} {\bibinfo {author} {\bibfnamefont {R.~Z.}\ \bibnamefont
  {Khaliullin}}, \bibinfo {author} {\bibfnamefont {M.}~\bibnamefont
  {Head-Gordon}}, \ and\ \bibinfo {author} {\bibfnamefont {A.~T.}\ \bibnamefont
  {Bell}},\ }\bibfield  {title} {\enquote {\bibinfo {title} {An efficient
  self-consistent field method for large systems of weakly interacting
  components},}\ }\href@noop {} {\bibfield  {journal} {\bibinfo  {journal} {J.
  Chem. Phys.},\ }\textbf {\bibinfo {volume} {124}},\ \bibinfo {pages} {204105}
  (\bibinfo {year} {2006})}\BibitemShut {NoStop}%
\bibitem [{\citenamefont {Vandevondele}\ \emph {et~al.}(2005)\citenamefont
  {Vandevondele}, \citenamefont {Krack}, \citenamefont {Mohamed}, \citenamefont
  {Parrinello}, \citenamefont {Chassaing},\ and\ \citenamefont
  {Hutter}}]{a:quickstep}%
  \BibitemOpen
  \bibfield  {author} {\bibinfo {author} {\bibfnamefont {J.}~\bibnamefont
  {Vandevondele}}, \bibinfo {author} {\bibfnamefont {M.}~\bibnamefont {Krack}},
  \bibinfo {author} {\bibfnamefont {F.}~\bibnamefont {Mohamed}}, \bibinfo
  {author} {\bibfnamefont {M.}~\bibnamefont {Parrinello}}, \bibinfo {author}
  {\bibfnamefont {T.}~\bibnamefont {Chassaing}}, \ and\ \bibinfo {author}
  {\bibfnamefont {J.}~\bibnamefont {Hutter}},\ }\bibfield  {title} {\enquote
  {\bibinfo {title} {Quickstep: Fast and accurate density functional
  calculations using a mixed gaussian and plane waves approach},}\ }\href@noop
  {} {\bibfield  {journal} {\bibinfo  {journal} {Comput. Phys. Commun.},\
  }\textbf {\bibinfo {volume} {167}},\ \bibinfo {pages} {103--128} (\bibinfo
  {year} {2005})}\BibitemShut {NoStop}%
\bibitem [{\citenamefont {Vandevondele}\ and\ \citenamefont
  {Hutter}(2007)}]{a:molopt}%
  \BibitemOpen
  \bibfield  {author} {\bibinfo {author} {\bibfnamefont {J.}~\bibnamefont
  {Vandevondele}}\ and\ \bibinfo {author} {\bibfnamefont {J.}~\bibnamefont
  {Hutter}},\ }\bibfield  {title} {\enquote {\bibinfo {title} {Gaussian basis
  sets for accurate calculations on molecular systems in gas and condensed
  phases},}\ }\href@noop {} {\bibfield  {journal} {\bibinfo  {journal} {J.
  Chem. Phys.},\ }\textbf {\bibinfo {volume} {127}},\ \bibinfo {pages} {114105}
  (\bibinfo {year} {2007})}\BibitemShut {NoStop}%
\bibitem [{\citenamefont {Hartwigsen}\ \emph {et~al.}(1998)\citenamefont
  {Hartwigsen}, \citenamefont {Goedecker},\ and\ \citenamefont
  {Hutter}}]{a:hgh}%
  \BibitemOpen
  \bibfield  {author} {\bibinfo {author} {\bibfnamefont {C.}~\bibnamefont
  {Hartwigsen}}, \bibinfo {author} {\bibfnamefont {S.}~\bibnamefont
  {Goedecker}}, \ and\ \bibinfo {author} {\bibfnamefont {J.}~\bibnamefont
  {Hutter}},\ }\bibfield  {title} {\enquote {\bibinfo {title} {Relativistic
  separable dual-space {G}aussian pseudopotentials from {H} to {Rn}},}\
  }\href@noop {} {\bibfield  {journal} {\bibinfo  {journal} {Phys. Rev. B},\
  }\textbf {\bibinfo {volume} {58}},\ \bibinfo {pages} {3641--3662} (\bibinfo
  {year} {1998})}\BibitemShut {NoStop}%
\end{thebibliography}

%

\end{document}